\PassOptionsToPackage{unicode}{hyperref}
\PassOptionsToPackage{hyphens}{url}
\documentclass[
]{article}
\usepackage{lmodern}
\usepackage{amssymb,amsmath}
\usepackage{ifxetex,ifluatex}
\ifnum 0\ifxetex 1\fi\ifluatex 1\fi=0 
  \usepackage[T1]{fontenc}
  \usepackage[utf8]{inputenc}
  \usepackage{textcomp} 
\else 
  \usepackage{unicode-math}
  \defaultfontfeatures{Scale=MatchLowercase}
  \defaultfontfeatures[\rmfamily]{Ligatures=TeX,Scale=1}
\fi
\IfFileExists{upquote.sty}{\usepackage{upquote}}{}
\IfFileExists{microtype.sty}{
  \usepackage[]{microtype}
  \UseMicrotypeSet[protrusion]{basicmath} 
}{}
\makeatletter
\@ifundefined{KOMAClassName}{
  \IfFileExists{parskip.sty}{%
    \usepackage{parskip}
  }{
    \setlength{\parindent}{0pt}
    \setlength{\parskip}{6pt plus 2pt minus 1pt}}
}{
  \KOMAoptions{parskip=half}}
\makeatother
\usepackage{xcolor}
\IfFileExists{xurl.sty}{\usepackage{xurl}}{} 
\IfFileExists{bookmark.sty}{\usepackage{bookmark}}{\usepackage{hyperref}}
\hypersetup{
  pdftitle={Comment on Gouriéroux, Monfort, Renne (2019): Identification and Estimation in Non-Fundamental Structural VARMA Models},
  pdfauthor={Bernd Funovits University of Helsinki and TU Wien},
  hidelinks,
  pdfcreator={LaTeX via pandoc}}
\usepackage[margin=1in]{geometry}
\usepackage{color}
\usepackage{fancyvrb}

\DefineVerbatimEnvironment{Highlighting}{Verbatim}{commandchars=\\\{\}}
\usepackage{framed}
\definecolor{shadecolor}{RGB}{248,248,248}
\newenvironment{Shaded}{\begin{snugshade}}{\end{snugshade}}

\newcommand{\CharTok}[1]{\textcolor[rgb]{0.31,0.60,0.02}{#1}}
\newcommand{\CommentTok}[1]{\textcolor[rgb]{0.56,0.35,0.01}{\textit{#1}}}

\newcommand{\ControlFlowTok}[1]{\textcolor[rgb]{0.13,0.29,0.53}{\textbf{#1}}}
\newcommand{\DataTypeTok}[1]{\textcolor[rgb]{0.13,0.29,0.53}{#1}}
\newcommand{\DecValTok}[1]{\textcolor[rgb]{0.00,0.00,0.81}{#1}}

\newcommand{\KeywordTok}[1]{\textcolor[rgb]{0.13,0.29,0.53}{\textbf{#1}}}
\newcommand{\NormalTok}[1]{#1}
\newcommand{\OperatorTok}[1]{\textcolor[rgb]{0.81,0.36,0.00}{\textbf{#1}}}
\newcommand{\OtherTok}[1]{\textcolor[rgb]{0.56,0.35,0.01}{#1}}

\newcommand{\StringTok}[1]{\textcolor[rgb]{0.31,0.60,0.02}{#1}}

\usepackage{graphicx}
\makeatletter
\def\maxwidth{\ifdim\Gin@nat@width>\linewidth\linewidth\else\Gin@nat@width\fi}
\def\maxheight{\ifdim\Gin@nat@height>\textheight\textheight\else\Gin@nat@height\fi}
\makeatother
\setkeys{Gin}{width=\maxwidth,height=\maxheight,keepaspectratio}
\makeatletter
\def\fps@figure{htbp}
\makeatother
\providecommand{\tightlist}{%
  \setlength{\itemsep}{0pt}\setlength{\parskip}{0pt}}
\usepackage{setspace}
\newlength{\cslhangindent}
\newenvironment{cslreferences}%
  {\setlength{\parindent}{0pt}%
  \everypar{\setlength{\hangindent}{\cslhangindent}}\ignorespaces}%
  {\par}

\title{Comment on Gouriéroux, Monfort, Renne (2019): Identification and
Estimation in Non-Fundamental Structural VARMA Models}
\author{Bernd Funovits \cr University of Helsinki and TU Wien}
\date{28 September 2020}

\begin{document}
\maketitle
\begin{abstract}
This comment points out a serious flaw in the article `Gouriéroux,
Monfort, Renne (2019): Identification and Estimation in Non-Fundamental
Structural VARMA Models', henceforth abbreviated as GMR, with regard to
mirroring complex-valued roots with Blaschke polynomial matrices.
Moreover, the (non-) feasibility of the proposed method (if the handling
of Blaschke transformation were not prohibitive) for cross-sectional
dimensions greater than two and vector moving average (VMA) polynomial
matrices of degree \(q\) greater than one is discussed.
\end{abstract}

\hypertarget{introduction}{%
\section{Introduction}\label{introduction}}

Blaschke matrices are at the center of GMR's estimation strategy.
Whenever the VMA polynomial matrix has complex valued determinantal
roots, GMR leaves the specified (real) parameter space and result in
complex-valued estimates. While the VMA matrix polynomial in GMR's
empirical application indeed has complex-valued determinantal roots, the
complex valued nature of the intermediate estimates does not propagate
to the final (published) estimates because they are ``covered up'' at
multiple locations. GMR discard the imaginary part whenever they result
in complex-valued parameter matrices or, if some quantity (similar to
the condition number of some ``Blaschke-transformed'' matrix) is
sufficiently large, replace a complex-valued matrix with the identity
matrix.\footnote{ Another example is that the imaginary part of the
  log-likelihood function is discarded.} It thus seems likely that the
authors were aware of the fact that their Blaschke matrices are
incorrectly complex-valued (which implies that the parameter space is
left) and that they have not dealt with the problem appropriately. Of
course, the obtained (mirrored) points in the parameter space do not
correspond to observationally equivalent models in terms of second
moments. Moreover, even the number of determinantal roots inside and
outside the unit circle (GMR call this ``fundamentalness regime'') may
be incorrect.

To fix ideas, consider the univariate MA(2) model
\(y_t = \left( 1 - \frac{3}{4} L \right) \left( 1 - \frac{1}{4} L \right) \varepsilon_t = \Theta(L) \varepsilon_t\)
where \(\left( \varepsilon_t \right)\) is non-Gaussian i.i.d. with unit
variance and \(L\) is the lag-operator such that
\(L \left(y_t\right) = \left(y_{t-1}\right)\). Blaschke factors of the
form \(g_1(z) = \frac{1-4z}{-4+z}\),
\(g_2(z) = \frac{1-\frac{4}{3} z}{-\frac{4}{3}+z}\),
\(g_3(z) = \frac{1-\frac{4}{3} z}{-\frac{4}{3}+z} \frac{1-4z}{-4+z}\),
for which \(g_i(z) g_i\left(\frac{1}{z}\right) = 1\) holds and which
mirror roots at the unit circle, can be used to obtain different
representations of \(\left(y_t\right)\) with identical second order
properties but different higher order properties, i.e.~the processes
\(\left( \Theta(L) g_i(L) \right) \left( g_i(L)^{-1} \varepsilon_t \right) = \Theta^{(i)}(L) \varepsilon^{(i)}_t\)
are indistinguishable from \(\left(y_t\right)\) in terms of second
moments. GMR's approach comprises obtaining all representations which
are observationally equivalent in terms of second moments and subsequent
optimisation in terms of their objective function (which uses more than
second order information) starting from all different combinations of
determinantal zeros inside or outside of the unit circle. While this is
easy in the univariate case with real roots, there are some non-trivial
problems in the multivariate case with complex determinantal roots of
the VMA matrix polynomial \(\Theta(z)\). Mirroring complex-conjugated
roots separately results necessarily in complex-valued parameter
matrices. Even when mirroring both complex-conjugated roots together,
the resulting VMA matrix polynomial is in general not real-valued and it
is non-trivial to ensure that the VMA matrix polynomial is real-valued
after mirroring both complex-conjugated roots (meaning that an
additional step is required).

Another issue, which does not invalidate GMR's approach as does the
first issue but still warrants mentioning, is that the method is
presented as if it were feasible for VARMA(p,q) models of arbitrary
polynomial degree \(q\) and cross-sectional dimension \(n\). This is not
the case. The article should at best be understood as a proposal of a
method for estimating VARMA(p,1) models of small dimensions (given a
correct handling of mirroring determinantal roots such that the
transformed system generates the same second moments as the original
one). Likely, it is not a coincidence that applications and examples
consider at most a cross-sectional dimension of two. In the code it is
even explicitly stated that it only works for \(q=1\). The computational
complexity of calculating all Blaschke matrices would be prohibitively
expensive when estimating models with a reasonable number of VMA lags or
cross-sectional dimension because for each choice of determinantal zeros
of the VMA polynomial matrix \(\Theta(z)\) inside or outside the unit
circle an optimisation has to be performed. If there are \(n \cdot q\)
real determinantal zeros, there are up to \(2^{nq}\) options, which is
prohibitive even for modest dimensions and lags. For \(4\) variables and
VMA order \(8\) (or similar), we would result in a maximal number of
\(2^{4\cdot8}\) representations (which are observationally equivalent in
terms of second moments but have different combinations of zeros inside
and outside the unit circle).\footnote{ Of course, one may only mirror
  pairs of complex-conjugated roots jointly in order not to leave the
  (real-valued) model class, which reduces this number. However, not
  even this fact is recognized in GMR. They mirror complex-conjugated
  roots separately.} If one optimisation based on a starting value of
this kind were to take one second (GMR's procedure is more costly), this
would result in a computing time of
\(2^{32} / \left(3600 \cdot 24 \cdot 365 \right) \approx 136\) years.
The fact that GMR's method is essentially only applicable for \(q=1\),
is alluded to shortly in Section \emph{4.1.4 Extension to the
SSVARMA(p,q) case} (which comprises 8 lines)\footnote{ In these 8 lines
  is another trivial error. GMR write that the ``eigenvalues of
  \(\tilde{\Theta}\) are the roots of \(\det\left(\Theta(z)\right)\)''.
  However, the eigenvalues of \(\tilde{\Theta}\) are the solutions of
  \(\det\left(\Theta\left(z^q\frac{1}{z}\right)\right)\) and (except for
  zeros at zero or infinity) correspond to the inverses of the
  determinantal roots of \(\Theta(z)\), see e.g.~\emph{Hannan, Deistler
  (2012, Theorem 1.2.2, page 19)}}, where the authors do mention that
their representation (4.18) is not the same as a VARMA(p,q).

The remainder of this comment is structured as follows: In Section 2, we
summarise the use of Blaschke transformations of VMA(1) polynomial
matrices (their method does not allow for VMA orders \(q>1\)) in GMR and
point out shortcomings. In Section 3, we provide real-valued
constructions of Blaschke polynomial matrices (for arbitrary VMA order
\(q\)). In Section 4, we illustrate and parametrise the extent to which
the use of Blaschke matrices in GMR is incorrect with various examples.
Lastly, we discuss GMR's estimation strategy and how the fact that the
real-valued model class is left is covered up when complex-valued
Blaschke polynomial matrices are post-multiplied on the VMA(1)
polynomial matrices.

\hypertarget{gmrs-incorrect-blaschke-procedure}{%
\section{GMR's Incorrect Blaschke
Procedure}\label{gmrs-incorrect-blaschke-procedure}}

GMR's parametrization of the VMA polynomial matrix is of the form
\(\Theta(L) = \left( I_n - \Theta L \right) C\). Their Blaschke
procedure for obtaining observationally equivalent VMA polynomial
matrices involves the following three functions:

\begin{itemize}
\tightlist
\item
  \emph{build.orthonormal.basis(\(\ldots\))}: It takes an eigenvector
  (pertaining to the determinantal root to be mirrored) of the VMA(1)
  coefficient matrix \(\Theta\) as input, ``normalises'' it (in an
  incorrect way), and attempts to build an orthonormal basis in an
  iterative process by projecting a basis vector on the orthogonal
  complement of orthonormal vectors from the previous iterative step.
  The following problems occur at this stage:

  \begin{itemize}
  \tightlist
  \item
    ``Normalising'' does not take the possibly complex nature of
    eigenvectors into account. According to this logic, the vector
    \(\left( \begin{smallmatrix} 1 \\ i \end{smallmatrix} \right)\)
    would have length zero.
  \item
    Multiplicity of eigenvalues greater than one are ignored (which is a
    minor problem because generically the multiplicity of eigenvalues is
    equal to one).
  \end{itemize}
\item
  \emph{compute.other.basic.form(\(\ldots\))}: Given the original VMA(1)
  coefficient matrix \(\Theta\), the original static shock transmission
  matrix \(C\), and a vector \(w\) of length \(n\) indicating whether a
  certain determinantal zero of \(\Theta(z)\) should be mirrored at the
  unit circle, this function computes an VMA matrix polynomial
  \(\Theta^{(1)}(L) = \left( I_n - \Theta^{(1)} L \right) C^{(1)}\)
  which generates the same spectral density (given a vector mutually and
  temporally vector of inputs with variance one) as the original
  \(\Theta(z)\) but whose determinantal zeros are mirrored at the unit
  circle according to the vector\footnote{The \(i\)-th zero is not
    mirrored if the \(i\)-th element of \(w\) is equal to \(1\).} \(w\).
  The vector \(w\) is the eigenvector pertaining to the determinantal
  zero of the VMA polynomial matrix to be mirrored. It serves as input
  to \emph{build.orthonormal.basis(\(\ldots\))}
\item
  \emph{compute.all.forms(\(\ldots\))}: This function is a wrapper which
  calls \emph{compute.other.basic.form(\(\ldots\))} for all possible
  combinations of zeros and ones in \(w\). It generates a matrix of
  dimension \(\left( 2^n \times n \right)\) where each row corresponds
  to a particular selection of determinantal zeros of \(\Theta(z)\) to
  be mirrored. In particular, complex-conjugated roots are mirrored
  separately, which results necessarily in complex-valued parameter
  matrices!\footnote{In order to see this, let us represent the zeros
    \(\alpha_{\pm}\) in
    \(\left(z-\alpha_{-}\right)\left(z-\alpha_{+}\right)\) in polar
    representation \(\alpha_{\pm}=r\cdot e^{\pm i\phi}\). Then, \[
    \begin{array}{rl}
    \left(z-\alpha_{-}\right)\left(z-\frac{1}{\overline{\alpha _{+}}}\right) &= \left(z-re^{-i\phi}\right)\left(z-\frac{1}{r}e^{i\phi}\right)\\
     &=z^{2}-z\left(re^{-i\phi}+\frac{1}{r}e^{i\phi}\right)+1
    \end{array}
    \] which is real only for \(\phi=k\cdot\pi,\ k\in\mathbb{Z}\),
    i.e.~when the root has trivial imaginary part, or when \(r=1\),
    which is excluded by the assumption of no determinantal roots on the
    unit circle.} Obviously, the dimension of this matrix makes GMR's
  approach prohibitively costly for VMA(q) processes with output
  dimension \(n > 2\) and even moderately high \(q>1\) (which is
  excluded in GMR's code).\footnote{ For example, consider a
    7-dimensional VMA(q) model with 4 lags which would result in
    268,435,456 starting values for their optimisation and a matrix
    requiring about 60 GB of memory.}
\end{itemize}

For more detail regarding versions of these functions, where variable
names (but no functionality) have been changed in order to increase
readability and comments have been added for clarification, we refer to
the Appendix.

The incorrect normalisation in
\emph{build.orthonormal.basis(\(\ldots\))}, which does not take the
complex-valued nature of eigenvectors into account, could be considered
an implementation error (albeit with serious implications). While the
facts that normalisation is done incorrectly and that complex-conjugated
roots are mirrored separately frankly add insult to injury, they are not
the focus of this comment. Much rather, this comment criticizes that GMR
ignored (possibly intentionally) a difficult theoretical problem: The
VMA matrix polynomial is in general not real-valued even when
complex-conjugated roots are mirrored jointly.\footnote{ This point is
  not addressed in the proofs of the spectral factorization theorem in
  \emph{Rozanov (1967)} and \emph{Hannan (1970)}, which use Blaschke
  matrices to mirror zeros of the VMA polynomial matrix outside the unit
  circle, either. Of course, this is not a problem in these books
  because the obtained spectral factors (which are in general not
  real-valued) do not serve as estimators but are rather an intermediate
  step in their proofs.} Without further adjustments, the parameter
space is thus left and the obtained intermediate estimators are
non-sensical.

In the following, it is shown that it is indeed possible to obtain
real-valued VMA polynomial matrices with mirrored complex-conjugated
roots but otherwise the same second order properties. This can be
achieved either with an additional unitary transformation after flipping
both complex-conjugated roots or with a state space construction.

\hypertarget{real-valued-approaches}{%
\section{Real-Valued Approaches}\label{real-valued-approaches}}

Here, we present correct real-valued approaches for applying all-pass
transformations (to be defined below) to a polynomial matrix
\(\Theta(z)\) of dimension \((n \times n)\) and degree \(q\) such that
the transformed polynomial matrix is again part of the model class (and
thus has real-valued coefficients). Like GMR, we assume that
\(\Theta(z)\) is non-singular at zero and non-singular on the unit
circle\footnote{For notational simplicity, we incorporate the static
  shock transmission matrix \(C\) into
  \(\Theta(z) = \Theta_0 + \Theta_1 z + \cdots + \Theta_q z^q\).}.

First, we provide some minimal definitions for all-pass transformations.
Subsequently, we discuss an approach\footnote{ Presented at the 12th
  International Conference on Computational and Financial Econometrics
  in Pisa in December 2018.} which involves the singular value
decomposition (SVD)\footnote{ Using the SVD has the advantage that it is
  numerically more stable and that it directly generates an orthogonal
  (or unitary) transformation (which is obviously all-pass). The
  orthogonalization step in GMR (starting from a normalised eigenvector
  and obtaining the remaining orthonormal basis vectors iteratively as
  orthogonal complements of the projection on the orthonormal vectors of
  the previous step) seems computationally costly and likely worse
  conditioned than the orthogonal or unitary matrices occurring in the
  SVD. Assuming of course a correct implementation, which is not the
  case for the code provided on the website of the Review of Economic
  Studies.} instead of the eigendecomposition in GMR. In contrast to
GMR, there is an additional step after mirroring a pair of
complex-conjugated roots of \(\det\left(\Theta(z)\right)\) which ensures
that the transformed matrix polynomial has real coefficients and is thus
still part of the model class under investigation. While this approach
is quite straight-forward, it is difficult to prove formally that the
coefficients of the transformed matrix polynomial are indeed real.

Lastly, we describe an approach based on state space methods which does
not take a detour into the complex plane and thus results by
construction in an all-pass filter with real coefficients. For details,
see \emph{Scherrer, Funovits (2020)}. Intermediate results (in
connection to the QR decomposition of the complex-conjugated
right-singular vectors) can be used to show that the coefficients in the
polynomial approach are real.

\hypertarget{definitions}{%
\subsection{Definitions}\label{definitions}}

A \emph{multivariate rational all-pass filter} is an
\((n\times n)\)-dimensional matrix \(V(z)\) whose entries are rational
functions and which satisfies
\(V(z)V^{*}\left(\frac{1}{z}\right)=V^{*}\left(\frac{1}{z}\right)V(z)=I_{n}\).
The superscript asterisk takes an (arbitrary) matrix function
\(m(z)=\sum_{j=-\infty}^{\infty}m_{j}z^{j}\) to its version with complex
conjugated and transposed coefficient matrices,
i.e.~\(m^{*}(z)=\sum_{j=-\infty}^{\infty}m_{j}^{*}z^{j}\).

Obviously, the spectral densities of \(y_t = \Theta(z) \eta_t\) and
\(y_t = \left(\Theta(z) V(z)\right) \left( V^*\left(\frac{1}{z}\right) \eta_t \right) = \tilde{\Theta}(z) \tilde{\eta}_t\),
where \(\eta_t\) is serially and mutually i.i.d. as in GMR, are
identical.

An \emph{elementary Blaschke factor} at \(\alpha\) (which is obviously
all-pass) is of the form\footnote{ Sometimes, the Blaschke factor is
  defined with an additional factor \(\alpha/ |\alpha|\). However, this
  factor is not well defined if \(\alpha = 0\).}

\[
B(z, \alpha) = \frac{1-\bar{\alpha} z}{-\alpha + z}.
\]

A \emph{squared Blaschke factor} for the pair of complex-conjugated
roots \(\alpha_{\pm}=\alpha_r \pm i\alpha_i\) (in obvious notation) is
defined as \[
B_{sq}(z, \alpha_{\pm}) = \frac{1-\bar{\alpha} z}{-\alpha + z} \frac{1-\alpha z}{-\bar{\alpha} + z} = \frac{1-2\alpha_r z+|\alpha|^2 z^2}{|\alpha|^2-2\alpha_r z + z^2}.
\] Lastly, a \emph{bivariate Blaschke factor} pertaining to the pair of
complex-conjugated roots \(\alpha_{\pm}=\alpha_r \pm i\alpha_i\), where
\(\alpha_i>0\), and the non-zero vector \(w \in \mathbb{C}^{2\times 1}\)
is given as \[
B_2(z, \alpha_{\pm}, w) = a^{-1}(z, \alpha_{\pm}) b(z, \alpha_{\pm}, w),
\] where \(a(z, \alpha_{\pm})\) is a diagonal matrix with entries
\(B_{sq}(z, \alpha_{\pm})\), and \(b(z, \alpha_{\pm}, w)\) is a
\((2 \times 2)\) polynomial matrix with highest degree \(2\) and which
is of reduced rank at \(z=\alpha^{-1}\), \(z=\bar\alpha^{-1}\),
\(z=\alpha\) and \(z=\bar\alpha\). We construct
\(b(z, \alpha_{\pm}, w)\) such that the column space of
\(b(\alpha_+, \alpha_{\pm}, w)\) is spanned by a given (non-trivial)
vector \(w \in \mathbb{C}^{2 \times 1}\).

\hypertarget{svd-approach-for-constructing-all-pass-filters}{%
\subsection{SVD Approach for Constructing All-Pass
Filters}\label{svd-approach-for-constructing-all-pass-filters}}

Irrespective of whether or not the root \(\alpha\) to be mirrored is
real or complex, we perform an SVD on \(\Theta(\alpha) = U D V^{*}\),
where \(U\) and \(V\) are orthogonal or unitary matrices, where the
asterisk denotes transposition and complex conjugation, and where \(D\)
is a diagonal matrix of non-negative and non-increasing elements
containing the singular values. For simplicity of presentation, we
assume that the rank deficiency of \(\Theta(\alpha)\) is equal to one
and thus that \(d_{nn}\) is equal to zero (and \(d_{n-1,n-1}\) is
positive).

Post-multiplying \(V\) on \(\Theta(\alpha)\) results in a matrix whose
Lastly column is zero. This implies that each element in the Lastly
column of \(\Theta(z) V\) has as polynomial factor \((z-\alpha)\).
Therefore, post-multiplying \(\mbox{diag}(I_{n-1}, B(z, \alpha))\)
mirrors the zero of \(\Theta(z)\) at \(\alpha\) to a zero at
\(\frac{1}{\bar{\alpha}}\). Obviously, orthogonal (or unitary)
transformations and transformations involving Blaschke factors are
all-pass.

In the case that \(\alpha_+\) is complex with positive imaginary part,
we perform the same procedure additionally on
\(\Theta(z) V_{+} \mbox{diag}(I_{n-1}, B(z, \alpha_+))\) such that also
the root with negative imaginary part, say \(\alpha_-\), is mirrored at
the unit circle. We obtain
\(\tilde{\Theta}(z) = \Theta(z) V_{+} \mbox{diag}(I_{n-1}, B(z, \alpha_+)) V_{-} \mbox{diag}(I_{n-1}, B(z, \alpha_{-}))\)
where the unitary matrix \(V_{-}\) is obtained from the SVD of
\(\Theta(\alpha_{-}) V_{+} \mbox{diag}(I_{n-1}, B(\alpha_{-}, \alpha_+))\).
Otherwise (if only one determinantal root from the pair of complex
conjugated roots is mirrored into the unit circle), it is impossible to
obtain a real-valued polynomial matrix. Be that as it may, in GMR,
complex-conjugated roots are reflected separately and therefore the
model class is necessarily left.

So far, we have used two static, in general complex-valued, unitary
transformations \(V_{+}\) and \(V_{-}\), and two dynamic (rational)
transformations involving \(B(z, \alpha_+)\) and \(B(z, \alpha_-)\)
whose product has real-valued coefficients. This suggests that another
static transformation may take us back into the model class with
real-valued parameter matrices. We construct a unitary transformation by
using the fact that
\(\tilde{\Theta}(1) \left(\tilde{\Theta}(1)\right)^* = \Theta(1) \left(\Theta(1)\right)^*\)
is real-valued together with the polar decomposition of
\(\tilde{\Theta}(1)\). Indeed, the real-valued orthogonal eigenbasis of
\(\Theta(1) \left(\Theta(1)\right)^*\) corresponds to the left singular
vectors of \(\Theta(1)\). Given the SVD \(\tilde{\Theta}(1) = W S X^*\),
its polar decomposition is
\(\tilde{\Theta}(1) = \left(W D W' \right) \left( W X^* \right)\). It
follows that right-multiplying the unitary matrix \(X W'\) results in
\(\tilde{\Theta}(1) \left(X W'\right)\) being real.

While this argument seems compelling, and while the transformations are
obviously all-pass and the implementation based on these derivations has
turned out to result in real-valued parameter matrices, it is difficult
to prove that \(\Theta(z)\) is real-valued whenever \(z\in\mathbb{R}\)
and \(z\neq 1\) by analysing the specific SVDs and the polar
decomposition of the matrices involved.

Therefore, we outline a different construction (for more detail see
\emph{Scherrer, Funovits (2020)}) which involves the bivariate Blaschke
factor \(B_2\left(z, \alpha_{\pm}, w \right)\) and which does not take a
detour into complex-valued matrices (implying that the model class is
left temporarily). A key step in this construction is the QR
decomposition of the real and imaginary part of the (complex-valued)
vector in the right-kernel of \(\Theta\left(\alpha_+\right)\). It is
possible to use the upper triangular matrix \(R\) from this QR
decomposition to parametrise the unitary matrices involved in the SVD
construction above and prove that \(\Theta(z)\) is indeed real-valued
whenever \(z\in\mathbb{R}\).

\hypertarget{proving-realness}{%
\subsection{Proving Realness}\label{proving-realness}}

\hypertarget{qr-decomposition-of-complex-valued-right-kernel}{%
\subsubsection{QR Decomposition of Complex-Valued
Right-Kernel}\label{qr-decomposition-of-complex-valued-right-kernel}}

We start from a (normalised) vector
\(v = v_r + i v_i \in \mathbb{C}^{n \times 1}\) in the right-kernel of
\(\Theta\left(\alpha_+\right)\), where
\(v_r, v_i \in \mathbb{R}^{n \times 1}\) are linearly independent.
Eventually, we will result in the transformed polynomial matrix (with
identical second order properties as the original one) \[
\Theta(z) \tilde{Q} \left[ \mbox{diag}\left\{B_2\left(z, \alpha_{\pm}, R \begin{pmatrix}1 \\ i\end{pmatrix}\right), I_{n-2}\right\} \right].
\] where the orthogonal real matrix \(\tilde{Q}\) and the
upper-triangular matrix \(R\) with positive diagonal elements are
obtained from the QR decomposition \[
\begin{pmatrix} 
v_r & v_i 
\end{pmatrix} 
= \tilde{Q} \tilde{R} 
= \begin{pmatrix}
Q & \tilde{Q}_2
\end{pmatrix} 
\begin{pmatrix}
R \\ 
0_{(n-2)\times 2} 
\end{pmatrix}.
\]

Note that \(R \left(\begin{smallmatrix}1 \\ i\end{smallmatrix}\right)\)
is in the right-kernel of \(\Theta\left(\alpha_+\right) Q\) and that, if
the column space of \(b\left(\alpha_+, \alpha_{\pm}, w \right)\) is
spanned by
\(w = R \left(\begin{smallmatrix}1 \\ i\end{smallmatrix}\right)\), we
have that
\(\Theta(\alpha_+) Q b\left(\alpha_+, \alpha_{\pm}, R \left(\begin{smallmatrix}1 \\ i\end{smallmatrix}\right)\right) = \Theta(\alpha_-) Q b\left(\alpha_-, \alpha_{\pm}, R \left(\begin{smallmatrix}1 \\ i\end{smallmatrix}\right)\right) = 0\).
Hence, all entries of
\(\Theta(z) Q b\left(\alpha_-, \alpha_{\pm}, R \left(\begin{smallmatrix}1 \\ i\end{smallmatrix}\right)\right)\)
and
\(\Theta(z) Q b\left(\alpha_+, \alpha_{\pm}, R \left(\begin{smallmatrix}1 \\ i\end{smallmatrix}\right)\right)\)
are divisible by the diagonal element of \(a(z, \alpha_{\pm})\).

The most elegant approach to constructing a polynomial matrix \(b(z)\)
with real coefficients is a state space construction. However, it is
also possible to parametrise unitary \((2\times2)\)-dimensional matrices
\(V_{\beta}\), \(V_{\gamma}\), and \(V_{\delta}\) with the parameters in
\(R\) and \(\alpha_+ = r\cdot e^{i\phi }\) such that \[
\tilde{Q} \cdot \mbox{diag}\left( V_{\beta}, I_{n-2}\right) \cdot \mbox{diag}\left( B(z, \alpha_+), I_{n-1} \right) \cdot \mbox{diag}\left( V_{\gamma}, I_{n-2}\right) \cdot \mbox{diag}\left( B(z, \alpha_-), I_{n-1}\right) \cdot \mbox{diag}\left( V_{\delta}, I_{n-2}\right)
\] is real. We will now give a succinct overview of both approaches and
refer to \emph{Scherrer, Funovits (2020)} for details.

\hypertarget{summary-of-state-space-construction}{%
\subsubsection{Summary of State Space
Construction}\label{summary-of-state-space-construction}}

The \((2 \times 2)\)-dimensional, real-valued, rational, all-pass filter
\(B_2(z, \alpha_{\pm}, w) = C(z^{-1}I_2 - A)^{-1} B + D\) is constructed
in the following steps.

First, the poles are fixed by setting
\(A = \left( \begin{smallmatrix} \lambda_r & \lambda_i \\ -\lambda_i & \lambda_r \end{smallmatrix} \right)\),
where \(\lambda_+ = \lambda_r + i\lambda_i = \alpha_+^{-1}\). Second,
\(C\) is determined by requiring that the column-space of
\(b(\alpha_+, \alpha_{\pm}, w)\) be equal to a vector
\(w \in \mathbb{C}^{2\times1}\) (and thus that the column-space of
\(b(\alpha_-, \alpha_{\pm}, w) = \overline{b(\alpha_+, \alpha_{\pm}, w)}\)
be equal to \(\bar{w} \in \mathbb{C}^{2\times1}\)). Lastly, \(B\) and
\(D\) are determined such that
\(B_2(z, \alpha_{\pm}, w) B_2'\left(\frac{1}{z}(z, \alpha_{\pm}, w\right)\)
is constant and that
\(B_2(z, \alpha_{\pm}, w) B_2'\left(\frac{1}{z}, \alpha_{\pm}, w\right) = I_n\).
For this Lastly step, it is necessary to apply a state transformation
and solve a Lyapunov equation.

\hypertarget{summary-of-parametrization-of-unitary-matrices}{%
\subsubsection{Summary of Parametrization of Unitary
Matrices}\label{summary-of-parametrization-of-unitary-matrices}}

Every unitary matrix can be parametrized through four parameters
\((\phi_0, \phi_1, \phi_2, \phi_3)\) as
\(e^{i \frac{\phi_0}{2}} \left( \begin{smallmatrix} e^{i \phi_1} \cos(\phi_3) & e^{i \phi_2} \sin(\phi_3) \\ -e^{-i \phi_2} \sin(\phi_3) & e^{-i \phi_1} \cos(\phi_3) \end{smallmatrix} \right)\).
Thus, we obtain \(V_{\beta}\) with \((\beta_0, \beta_1) = (0, 0)\) by
choosing\footnote{ The parameters \(\beta_2, \beta_3\) can be determined
  as a function of the parameters in \(R\).} \(\beta_2, \beta_3\) such
that \(R\left(\begin{smallmatrix} 1 \\ i\end{smallmatrix}\right)\) is in
the span of
\(\left(\begin{smallmatrix} \cos(\beta_3) e^{i \beta_2} \\ \sin(\beta_3) \end{smallmatrix}\right)\).
Similarly, \(V_{\gamma}\) with \((\gamma_0, \gamma_1) = (0, 0)\) is
determined by setting \(\gamma_2, \gamma_3\) such that
\(R\left(\begin{smallmatrix}1 \\ -i\end{smallmatrix}\right)\) is in the
span of
\(V_{\beta} \cdot \left( \begin{smallmatrix} B(\alpha_-,\alpha_+) & 0 \\ 0 & 1 \end{smallmatrix} \right) \cdot V_{\gamma,[\bullet,1]}\).
Note that the parameters \(\gamma_2, \gamma_3\) are functions of
\(\beta_2, \beta_3\), and \(\alpha_+\). Lastly, \(V_{\delta}\) is chosen
such that
\(V_{\beta} \cdot \left( \begin{smallmatrix} B(1,\alpha_+) & 0 \\ 0 & 1 \end{smallmatrix} \right) \cdot V_{\gamma} \cdot \left( \begin{smallmatrix} B(1,\alpha_-) & 0 \\ 0 & 1 \end{smallmatrix} \right) \cdot V_{\delta}\)
is equal to the identity matrix. Using this parametrisation, it can be
verified by straight-forward calculations that the obtained
transformation has real coefficients.

\hypertarget{examples}{%
\section{Examples}\label{examples}}

In this section, we parametrise matrices in order to analyze how the
construction in GMR fails. We start with the purely complex case, go on
to skew-symmetric VMA(1) coefficients, and finally consider some general
coefficient matrices which result in complex-valued determinantal zeros
of \(\Theta(z)\). We will consider the \(2\)-dimensional case, because
this is the one used in GMR. Code for these examples can be found in the
Appendix.

\hypertarget{incorrect-normalisation}{%
\subsection{Incorrect ``Normalisation''}\label{incorrect-normalisation}}

\hypertarget{purely-complex-case}{%
\subsubsection{Purely Complex Case}\label{purely-complex-case}}

Generating a VMA(1) coefficient matrix of the kind
\(\left(\begin{smallmatrix} 0 & b \\ -b & 0 \end{smallmatrix}\right)\)
(randomly) results in an error due to the fact that in GMR, the squared
length of a vector (which is subsequently used to ``normalise'' the
vector) is incorrectly calculated by summing the squared components.
Since the eigenvectors of these matrices are of the form
\(\left(\begin{smallmatrix} 1 \\ \pm i \end{smallmatrix}\right)\), the
implementation in GMR is non-functional because calculating the sum of
the squared components of this vector is zero.

\hypertarget{skew-symmetric-complex-case}{%
\subsubsection{Skew-Symmetric Complex
Case}\label{skew-symmetric-complex-case}}

Next, we consider VMA(1) coefficient matrices of the kind
\(\left(\begin{smallmatrix} a & b \\ -b & a \end{smallmatrix}\right)\).
In exact arithmetic, the eigenvectors are again of the form
\(\left(\begin{smallmatrix} 1 \\ \pm i \end{smallmatrix}\right)\). Since
GMR check whether the sum of squared components is
\emph{exactly}\footnote{More commonly, one checks whether a quantity is
  numerically zero by comparing it to a certain small threshold. A
  common way to check whether a quantity is numerically zero is to
  compare its absolute value to a certain numerical tolerance level.}
equal to zero, this sometimes does not result in an error but rather in
very large (and non-sensical) entries of the VMA(1) coefficient matrix
\(\Theta\) and the static shock transmission matrix \(C\). For example,
setting \(a=b=1\) results in an error (or rather a message printed to
the console and an error triggered at later stage) but setting \(a=2\)
and \(b=5\) results in very large entries and incorrect results (in the
sense that the obtained polynomial matrix is not all-pass and does not
have its roots mirrored).

\hypertarget{general-complex-case}{%
\subsubsection{General Complex Case}\label{general-complex-case}}

Whenever the VMA(1) coefficient matrix is not exactly skew-symmetric,
the procedure does not throw an error. However, depending on the extent
to which the matrix is not skew-symmetric, the results remain incorrect.
In order to understand this, one may consider matrices of the kind
\(\left(\begin{smallmatrix} a & b + c \\ -b & a \end{smallmatrix}\right)\),
where \(c \cdot b > 0\) such that we obtain complex-valued eigenvalues.
Depending on the absolute value of \(c\), it is possible to parametrise
the distance from transformations which are actually all-pass.

\hypertarget{fundamental-issues}{%
\subsection{Fundamental Issues}\label{fundamental-issues}}

Even without the problems mentioned above, the strategy in GMR would
still fail for two reasons. First, complex-conjugated roots \(\alpha_+\)
and \(\alpha_-\) are mirrored separately in GMR. This results in
complex-conjugated parameter matrices which cannot be made real and are
thus not part of the parameter space specified in GMR. Second, even when
both complex-conjugated roots are mirrored (and a real-valued
construction would be possible), the method used in GMR still results in
complex-valued parameter matrices \(\Theta\) and \(C\). Given that GMR
go to great lengths to deal with complex-valued quantities in their
code, it is hard to imagine that GMR were not aware of this shortcoming.
Discarding the imaginary part of these matrices results in parameters
which do not correspond to observationally equivalent models in terms of
second moments. In addition, the number of determinantal roots inside
the unit circle may change when discarding imaginary parts.

\hypertarget{issues-in-gmrs-estimation-strategy}{%
\section{Issues in GMR's Estimation
Strategy}\label{issues-in-gmrs-estimation-strategy}}

In this section, we discuss in detail where the questionable steps in
GMR's code (Supplementary data, downloaded from the website of The
Review of Economic Studies) occur. In particular, we describe where
complex-valued quantities are replaced with (ad-hoc choices of) real
ones. We emphasize that in every evaluation of the likelihood function,
an adjustment for the possibility of complex-valued matrices is made.
All line references refer to GMR's script \emph{set.of.procedures.R}.

The script \emph{run.VARMA.BQ.dataset.R} calls
\emph{run.estim.QZ.GMM.MLE.R} (in the global environment) with different
values for the AR order (1 to 6). The VMA order is always equal to one.
For each AR order, the latter script calls first the GMM method
\emph{estim.VARMAp1.2SLS.GMM(\(\ldots\))} (serving as initial estimates
of their maxmimum likelihood (ML) approach) and the the ML optimisation
routine \emph{estim.MA.inversion(\(\ldots\))}. These two main estimation
procedures (as well as the parameter choices in the two scripts
\emph{run.estim.QZ.GMM.MLE.R} and \emph{run.VARMA.BQ.dataset.R}) are
described in detail in the Appendix.

\hypertarget{issues-in-gmm-procedure}{%
\subsection{Issues in GMM Procedure}\label{issues-in-gmm-procedure}}

In the main function \emph{estim.VARMAp1.2SLS.GMM(\(\ldots\))} (after
computing ``all observationally equivalent'' VMA polynomial matrices in
line 1051), GMR apply \emph{Re(\(\cdot\))}, which returns the real part
of a complex number, to \emph{all.MA\$all.Theta} and
\emph{all.MA\$all.C} (which are the arrays which describing all putative
observationally equivalent VMA polynomial matrices). There is of course
no foundation for doing so.\footnote{ An issue unrelated to Blaschke
  polynomial matrices appears in line 1470 where the absolute value
  function \emph{abs(\(\cdot\))} is only applied to the diagonal
  elements the sign change should be applied to the whole column. This
  seems problematic because it happens directly before the variables are
  returned, i.e.~there is no optimisation which could alleviate wrong
  parameter values.}

\hypertarget{issues-in-ml-procedure}{%
\subsection{Issues in ML Procedure}\label{issues-in-ml-procedure}}

Similar to the GMM procedure, in the main function
\emph{estim.MA.inversion(\(\ldots\))} (after computing ``all
observationally equivalent'' VMA polynomials in line 274), GMR apply
\emph{Re(\(\cdot\))} to \emph{all.MA\$all.Theta} and
\emph{all.MA\$all.C}. Again, naturally, there is no foundation for doing
so.

An even more serious shortcoming appears in the function
\emph{estim.struct.shocks(\(\ldots\))} (which can be considered to be at
the core of the implementation in GMR since it uses the Schur
decomposition to ``invert'' the VMA polynomial), where the static shock
transmission matrix \(C\) is replaced with an identity matrix whenever a
quantity similar to the condition number\footnote{ The condition number
  of a matrix is usually defined as the ratio of the largest to the
  smallest singular value. GMR calculate the ratio of the largest to the
  smallest absolute value of the eigenvalues.} is above a certain large
threshold or any element of the vector containing the absolute value of
the ratios of the eigenvalues of the imaginary part of \(C\) to the real
part of \(C\) has a large element. This is, of course, an ad-hoc fix of
a non-functional procedure and readers are misled to believe that
Blaschke matrices are used even though the obtained values can, at best,
be interpreted as ad-hoc selected new starting values for the
optimisation of an objective function. Whenever none of the two
conditions above is satisfied, GMR continue with a complex-valued static
shock transmission matrix \(C\) and VMA(1) coefficient matrix
\(\Theta\). The reason why no complex-valued parameters appear at the
end of the procedure is found in various calls to \emph{Re(\(\cdot\))},
in order to discard imaginary parts (e.g.~in the likelihood evaluation
function \emph{ML.inversion.loglik(\(\ldots\))} on line 171).

Furthermore, we want to emphasize that
\emph{estim.struct.shocks(\(\ldots\))} is at the center of GMR's
implementation (which can also be seen from the dependency graphs in the
Appendix). Every time the likelihood is evaluated (e.g.~in the
optimisation routine \emph{stats::optim(\(\ldots\))}),
\emph{estim.struct.shocks(\(\ldots\))} is called by
\emph{ML.inversion.loglik.aux(\(\ldots\))}. The latter function calls
\emph{g(\(\ldots\))} (line 2401), which in turn calls the respective
class of log-densities (usually a mixture of Gaussians). There is an
additional ad-hoc fix\footnote{ This might be reasonable to avoid
  singular cases. Note however, that singular \(C\) matrices only occur
  because of an incorrect implementation of mirroring roots with
  Blaschke matrices.} in \emph{g(\(\ldots\))} (next to setting \(C\)
equal to the identity matrix if one of the two conditions mentioned
above is not satisfied): Whenever the sum of the absolute values of the
eigenvalues of \(C\) is small (i.e.~the matrix \(C\) is close to
singular), the value of the log density is set to \(-100000\). Similar
to \emph{estim.struct.shocks(\(\ldots\))}, the function
\emph{g(\(\ldots\))} is called whenever the likelihood is evaluated and
is therefore also at the center of the procedure in GMR.

\hypertarget{acknowledgments}{%
\section{Acknowledgments}\label{acknowledgments}}

Financial support by the Research Funds of the University of Helsinki as
well as by funds of the Oesterreichische Nationalbank (Austrian Central
Bank, Anniversary Fund, project number: 17646) is gratefully
acknowledged.

\hypertarget{references}{%
\section{References}\label{references}}

\hypertarget{refs}{}
\begin{cslreferences}
\leavevmode\hypertarget{ref-funovits2020identifiability}{}%
Funovits, Bernd. 2020. ``Identifiability and Estimation of Possibly
Non-Invertible Svarma Models: A New Parametrisation.''
\url{http://arxiv.org/abs/2002.04346}.

\leavevmode\hypertarget{ref-GourierouxMR_svarma19}{}%
Gouriéroux, Christian, Alain Monfort, and Jean-Paul Renne. 2019.
``Identification and Estimation in Non-Fundamental Structural VARMA
Models.'' \emph{Review of Economic Studies}, 1--39.
\url{https://doi.org/10.1093/restud/rdz028}.

\leavevmode\hypertarget{ref-Hannan70}{}%
Hannan, Edward J. 1970. \emph{Multiple Time Series}. Wiley.

\leavevmode\hypertarget{ref-HannanDeistler12}{}%
Hannan, Edward J., and Manfred Deistler. 2012. \emph{The Statistical
Theory of Linear Systems}. Philadelphia: SIAM Classics in Applied
Mathematics.

\leavevmode\hypertarget{ref-LanneSaikkonen13}{}%
Lanne, Markku, and Pentti Saikkonen. 2013. ``Noncausal Vector
Autoregression.'' \emph{Econometric Theory} 29 (03): 447--81.
\url{https://doi.org/10.1017/S0266466612000448}.

\leavevmode\hypertarget{ref-Rozanov67}{}%
Rozanov, Yuri A. 1967. \emph{Stationary Random Processes}. San
Francisco: Holden-Day.

\leavevmode\hypertarget{ref-ScherrerFuno20stsp_blaschke}{}%
Scherrer, Wolfgang, and Bernd Funovits. 2019. ``All-Pass Filters for
Mirroring Pairs of Complex-Conjugated Roots.''
\url{http://arxiv.org/abs/2010.01598}.

\leavevmode\hypertarget{ref-VelascoLobato18}{}%
Velasco, Carlos, and Ignacio N. Lobato. 2018. ``Frequency Domain Minimum
Distance Inference for Possibly Noninvertible and Noncausal Arma
Models.'' \emph{Ann. Statist.} 46 (2): 555--79.
\url{https://doi.org/10.1214/17-AOS1560}.
\end{cslreferences}

\newpage

\hypertarget{appendix}{%
\section{Appendix}\label{appendix}}

In this Appendix, we provide more details regarding GMR's implementation
of the Blaschke procedure (i.e.~we discuss GMR's code), provide code for
and run the examples mentioned in the main text, and describe the
estimation procedure in detail.

\hypertarget{gmrs-implementation-of-blaschke-matrices}{%
\section{GMR's Implementation of Blaschke
Matrices}\label{gmrs-implementation-of-blaschke-matrices}}

The functions below are taken from the website of the Review of Economic
Studies. Variable names have been changed in order to increase
readability and comments have been added for clarification. For testing
purposes, we have also included the original functions where no names
have been changed and no comments have been added

\begin{Shaded}
\begin{Highlighting}[]
\ControlFlowTok{if}\NormalTok{(params}\OperatorTok{$}\NormalTok{use\_GMRcode\_commented)\{}
  
\CommentTok{\#\textquotesingle{} Build Orthonormal Basis }
\CommentTok{\#\textquotesingle{} }
\CommentTok{\#\textquotesingle{} This function creates an orthonormal matrix \textbackslash{}code\{orth\} whose first column }
\CommentTok{\#\textquotesingle{} is the "normalised" value of \textbackslash{}code\{vec\}.}
\CommentTok{\#\textquotesingle{}}
\CommentTok{\#\textquotesingle{} @param vec Vector to be "normalised" of dimension \textbackslash{}code\{dim\_out\}}
\CommentTok{\#\textquotesingle{}}
\CommentTok{\#\textquotesingle{} @return Orthonormal matrix \textbackslash{}code\{K\} (no mentioning about the complex case, }
\CommentTok{\#\textquotesingle{} seems like it either fails or result is incorrect in the complex case)}
\CommentTok{\#\textquotesingle{} @export}
\NormalTok{build.orthonormal.basis \textless{}{-}}\StringTok{ }\ControlFlowTok{function}\NormalTok{(vec)\{}
  
  \CommentTok{\# Integer{-}valued parameters}
\NormalTok{  dim\_out \textless{}{-}}\StringTok{ }\KeywordTok{length}\NormalTok{(vec)}
  
  \CommentTok{\# Allocate other columns to be orthogonalized w.r.t. input *vec*}
\NormalTok{  Id \textless{}{-}}\StringTok{ }\KeywordTok{diag}\NormalTok{(dim\_out)}
  \ControlFlowTok{if}\NormalTok{ (vec[}\DecValTok{1}\NormalTok{]}\OperatorTok{!=}\DecValTok{0}\NormalTok{) \{}
\NormalTok{    basis\_non\_orth \textless{}{-}}\StringTok{ }\KeywordTok{matrix}\NormalTok{(Id[,}\DecValTok{2}\OperatorTok{:}\NormalTok{dim\_out], dim\_out, dim\_out}\DecValTok{{-}1}\NormalTok{)}
\NormalTok{  \} }\ControlFlowTok{else}\NormalTok{ \{}
\NormalTok{    basis\_non\_orth \textless{}{-}}\StringTok{ }\KeywordTok{matrix}\NormalTok{(Id[,}\DecValTok{1}\OperatorTok{:}\NormalTok{(dim\_out}\DecValTok{{-}1}\NormalTok{)], dim\_out, dim\_out}\DecValTok{{-}1}\NormalTok{)}
\NormalTok{  \}}
  
  \CommentTok{\# In the following if{-}condition, no thought is spent on the complex case.}
  \CommentTok{\# (sum(c(1, 1i)\^{}2) == 0 is TRUE)}
  \ControlFlowTok{if}\NormalTok{(}\KeywordTok{sum}\NormalTok{(vec}\OperatorTok{\^{}}\DecValTok{2}\NormalTok{) }\OperatorTok{==}\StringTok{ }\DecValTok{0}\NormalTok{)\{}
    
    \CommentTok{\# This should throw an error, not print a message.}
    \CommentTok{\# Of course, an error is thrown later in the estimation procedure  }
    \CommentTok{\# because of a dimension mismatch.}
    \KeywordTok{print}\NormalTok{(}\StringTok{"z should not be 0"}\NormalTok{)}
\NormalTok{    orth \textless{}{-}}\StringTok{ }\OtherTok{NaN}
    
\NormalTok{  \} }\ControlFlowTok{else}\NormalTok{ \{}
    
    \CommentTok{\# Normalise the column and initialize successive orthonormalisation}
\NormalTok{    orth \textless{}{-}}\StringTok{ }\KeywordTok{matrix}\NormalTok{(vec}\OperatorTok{/}\KeywordTok{sqrt}\NormalTok{(}\KeywordTok{sum}\NormalTok{(vec}\OperatorTok{\^{}}\DecValTok{2}\NormalTok{)), }\DataTypeTok{ncol =} \DecValTok{1}\NormalTok{)}
    
    \CommentTok{\# Successively orthonormalise columns of *basis\_non\_orth*}
    \ControlFlowTok{for}\NormalTok{(ix\_col }\ControlFlowTok{in} \DecValTok{2}\OperatorTok{:}\NormalTok{dim\_out)\{}
      
      \CommentTok{\# Columns that are already orthonormal}
\NormalTok{      X \textless{}{-}}\StringTok{ }\NormalTok{orth[, }\DecValTok{1}\OperatorTok{:}\NormalTok{(ix\_col}\DecValTok{{-}1}\NormalTok{)]}
      
      \CommentTok{\# Next column to be orthonormalised}
      \CommentTok{\# Note that *basis\_non\_orth* is a matrix with *dim\_out{-}1* columns!}
\NormalTok{      y \textless{}{-}}\StringTok{ }\KeywordTok{matrix}\NormalTok{(basis\_non\_orth[, ix\_col}\DecValTok{{-}1}\NormalTok{], }\DataTypeTok{ncol =} \DecValTok{1}\NormalTok{)}
      
      \CommentTok{\# New orthogonal column}
\NormalTok{      orth\_col\_new \textless{}{-}}\StringTok{ }\NormalTok{(Id }\OperatorTok{{-}}\StringTok{ }\NormalTok{X }\OperatorTok{\%*\%}\StringTok{ }\KeywordTok{solve}\NormalTok{(}\KeywordTok{t}\NormalTok{(X) }\OperatorTok{\%*\%}\StringTok{ }\NormalTok{X) }\OperatorTok{\%*\%}\StringTok{ }\KeywordTok{t}\NormalTok{(X)) }\OperatorTok{\%*\%}\StringTok{ }\NormalTok{y}
      
      \CommentTok{\# Normalise}
\NormalTok{      orth\_col\_new \textless{}{-}}\StringTok{ }\NormalTok{orth\_col\_new}\OperatorTok{/}\KeywordTok{sqrt}\NormalTok{(}\KeywordTok{sum}\NormalTok{(orth\_col\_new}\OperatorTok{\^{}}\DecValTok{2}\NormalTok{))}
      
      \CommentTok{\# Bind old and now orthonormal columns together}
\NormalTok{      orth \textless{}{-}}\StringTok{ }\KeywordTok{cbind}\NormalTok{(orth, orth\_col\_new)}
\NormalTok{    \}}
\NormalTok{  \}}
  
  \KeywordTok{return}\NormalTok{(orth)}
\NormalTok{\}}

\CommentTok{\#\textquotesingle{} GMR Function: Helper for computing Blaschke Transformations}
\CommentTok{\#\textquotesingle{} }
\CommentTok{\#\textquotesingle{} For given vector \textbackslash{}code\{w\} of dimension \textbackslash{}code\{dim\_out\}, calculate }
\CommentTok{\#\textquotesingle{} a new MA polynomial matrix and associated static shock transmission matrix.}
\CommentTok{\#\textquotesingle{} }
\CommentTok{\#\textquotesingle{} @section GMR "documentation":}
\CommentTok{\#\textquotesingle{} Entries are 1 or {-}1 (see Lippi and Reichlin 1994).}
\CommentTok{\#\textquotesingle{} If \textbackslash{}code\{w[i]={-}1\} then the output MA representation corresponds }
\CommentTok{\#\textquotesingle{} to an MA representation }
\CommentTok{\#\textquotesingle{} with the same spectral density as the input one, }
\CommentTok{\#\textquotesingle{} but the i{-}th pole of det(I {-} Theta.L) would have been replaced by }
\CommentTok{\#\textquotesingle{} its inverse conjugate.}
\CommentTok{\#\textquotesingle{} \textbackslash{}strong\{The condition is rather: }
\CommentTok{\#\textquotesingle{}         If \textbackslash{}code\{w[i] != 1\}, the zero gets flipped.\}}
\CommentTok{\#\textquotesingle{}}
\CommentTok{\#\textquotesingle{} @inheritParams compute.all.forms}
\CommentTok{\#\textquotesingle{} @param w Vector of length \textbackslash{}code\{dim\_out\}. }
\CommentTok{\#\textquotesingle{}   Indicates whether a certain zero is mirrored inside or not.}
\CommentTok{\#\textquotesingle{}}
\CommentTok{\#\textquotesingle{} @return List with slots}
\CommentTok{\#\textquotesingle{}         \textbackslash{}describe\{}
\CommentTok{\#\textquotesingle{}           \textbackslash{}item\{Theta: \}\{MA(1) coefficient matrix\}}
\CommentTok{\#\textquotesingle{}           \textbackslash{}item\{C: \}\{Static shock transmission matrix\}}
\CommentTok{\#\textquotesingle{}         \}}
\CommentTok{\#\textquotesingle{}         from which the new MA polynomial with flipped roots can be obtained.}
\CommentTok{\#\textquotesingle{}         }
\CommentTok{\#\textquotesingle{} @export}
\NormalTok{compute.other.basic.form \textless{}{-}}\StringTok{ }\ControlFlowTok{function}\NormalTok{(Theta, C, w)\{}
  
  \CommentTok{\# Integer{-}valued parameters}
\NormalTok{  n\_zeros \textless{}{-}}\StringTok{ }\KeywordTok{dim}\NormalTok{(C)[}\DecValTok{1}\NormalTok{]}
  
  \CommentTok{\# Only for real eigenvalues of Theta does this correspond to }
  \CommentTok{\# mirroring on the unit circle.}
  \CommentTok{\# For complex{-}valued ones this is (in polar coordinates): }
  \CommentTok{\# (r, e\^{}\{i s\}) =\textgreater{} (1/r, e\^{}\{{-}i s\})}
\NormalTok{  all\_zeros\_initial \textless{}{-}}\StringTok{ }\DecValTok{1}\OperatorTok{/}\KeywordTok{eigen}\NormalTok{(Theta)}\OperatorTok{$}\NormalTok{values}
  
\NormalTok{  Theta\_step \textless{}{-}}\StringTok{ }\NormalTok{Theta}
\NormalTok{  C\_step \textless{}{-}}\StringTok{ }\NormalTok{C}
  
  \CommentTok{\# Irrespective of whether the root at hand is complex or not}
  \ControlFlowTok{for}\NormalTok{(ix\_zero }\ControlFlowTok{in} \DecValTok{1}\OperatorTok{:}\NormalTok{n\_zeros)\{}
    
    \CommentTok{\# "Mirror" according to indicator  }
    \ControlFlowTok{if}\NormalTok{(w[ix\_zero] }\OperatorTok{!=}\StringTok{ }\DecValTok{1}\NormalTok{)\{}
      
      \CommentTok{\# Choose zero to be mirrored and get its position}
\NormalTok{      zero\_this \textless{}{-}}\StringTok{ }\NormalTok{all\_zeros\_initial[ix\_zero]}
\NormalTok{      all\_zeros\_step \textless{}{-}}\StringTok{ }\DecValTok{1}\OperatorTok{/}\KeywordTok{eigen}\NormalTok{(Theta\_step)}\OperatorTok{$}\NormalTok{values}
\NormalTok{      zero\_this\_position \textless{}{-}}\StringTok{ }
\StringTok{        }\KeywordTok{which}\NormalTok{(}\KeywordTok{abs}\NormalTok{(all\_zeros\_step}\OperatorTok{{-}}\NormalTok{zero\_this) }\OperatorTok{==}\StringTok{ }\KeywordTok{min}\NormalTok{(}\KeywordTok{abs}\NormalTok{(all\_zeros\_step}\OperatorTok{{-}}\NormalTok{zero\_this)))[}\DecValTok{1}\NormalTok{]}
      
      \CommentTok{\# Get the corresponding eigenvector and transform it with inverse of C}
\NormalTok{      ev\_theta\_step \textless{}{-}}\StringTok{ }\KeywordTok{eigen}\NormalTok{(Theta\_step)}\OperatorTok{$}\NormalTok{vectors[, zero\_this\_position]}
\NormalTok{      ev\_theta\_step\_transformed \textless{}{-}}\StringTok{ }\KeywordTok{solve}\NormalTok{(C\_step) }\OperatorTok{\%*\%}\StringTok{ }\NormalTok{ev\_theta\_step}
      
\NormalTok{      trafo\_orth \textless{}{-}}\StringTok{ }\KeywordTok{build.orthonormal.basis}\NormalTok{(ev\_theta\_step\_transformed)}
      
      \CommentTok{\# Transformation of the static shock transmission matrix C corresponds to:}
      \CommentTok{\# 1) Post{-}multiply orthogonal transformation}
      \CommentTok{\# 2) Post{-}multiply identity matrix with (1,1) element replaced }
      \CommentTok{\#    by 1/Conj(zero\_this)}
\NormalTok{      CK.mod \textless{}{-}}\StringTok{ }\NormalTok{C\_step }\OperatorTok{\%*\%}\StringTok{ }\NormalTok{trafo\_orth}
\NormalTok{      a \textless{}{-}}\StringTok{ }\NormalTok{CK.mod[, }\DecValTok{1}\NormalTok{]}
\NormalTok{      CK.mod[, }\DecValTok{1}\NormalTok{] \textless{}{-}}\StringTok{ }\OperatorTok{{-}}\StringTok{ }\NormalTok{a }\OperatorTok{*}\StringTok{ }\NormalTok{(}\DecValTok{1}\OperatorTok{/}\KeywordTok{Conj}\NormalTok{(zero\_this))}
      
      \CommentTok{\# Transformation of MA(1) coefficient corresponds to:}
      \CommentTok{\# 1) Post{-}multiply orthogonal transformation}
      \CommentTok{\# 2) Post{-}multiply identity matrix with (1,1) element replaced }
      \CommentTok{\#    by {-}zero\_this}
      \CommentTok{\# 3) Post{-}multiply newly transformed inverse of C\_step}
\NormalTok{      Theta.mod \textless{}{-}}\StringTok{ }\NormalTok{Theta\_step }\OperatorTok{\%*\%}\StringTok{ }\NormalTok{C\_step }\OperatorTok{\%*\%}\StringTok{ }\NormalTok{trafo\_orth}
\NormalTok{      Theta.mod[,}\DecValTok{1}\NormalTok{] \textless{}{-}}\StringTok{ }\OperatorTok{{-}}\StringTok{ }\NormalTok{a}
      
\NormalTok{      C\_step \textless{}{-}}\StringTok{ }\NormalTok{CK.mod}
\NormalTok{      Theta\_step \textless{}{-}}\StringTok{ }\NormalTok{Theta.mod }\OperatorTok{\%*\%}\StringTok{ }\KeywordTok{solve}\NormalTok{(CK.mod)}
\NormalTok{    \}}
\NormalTok{  \}}
  
  \KeywordTok{return}\NormalTok{(}\KeywordTok{list}\NormalTok{(}\DataTypeTok{Theta =}\NormalTok{ Theta\_step,}
              \DataTypeTok{C =}\NormalTok{ C\_step))}
\NormalTok{\}}

\CommentTok{\#\textquotesingle{} GMR Function: Obtain All Observationally Equivalent MA Polynomials}
\CommentTok{\#\textquotesingle{} }
\CommentTok{\#\textquotesingle{} Generates all combinations of zeros inside and outside the unit circle.}
\CommentTok{\#\textquotesingle{} }
\CommentTok{\#\textquotesingle{} This function is called by GMR\textquotesingle{}s \textbackslash{}code\{estim.MA.inversion\}.}
\CommentTok{\#\textquotesingle{}}
\CommentTok{\#\textquotesingle{} @param Theta MA(1) Parameter matrix}
\CommentTok{\#\textquotesingle{} @param C Static shock transmission matrix}
\CommentTok{\#\textquotesingle{}}
\CommentTok{\#\textquotesingle{} @return List with two slots, \textbackslash{}code\{all.Theta\} and \textbackslash{}code\{all.C\}. }
\CommentTok{\#\textquotesingle{}   Each slot contains an array of }
\CommentTok{\#\textquotesingle{}   dimension \textbackslash{}code\{(dim\_out, dim\_out, 2\^{}dim\_out)\}, }
\CommentTok{\#\textquotesingle{}   where the third dimension corresponds to the choice of }
\CommentTok{\#\textquotesingle{}   zeros inside and outside the unit circle.}
\CommentTok{\#\textquotesingle{} @export}
\NormalTok{compute.all.forms \textless{}{-}}\StringTok{ }\ControlFlowTok{function}\NormalTok{(Theta,C)\{}
  
\NormalTok{  dim\_out \textless{}{-}}\StringTok{ }\KeywordTok{dim}\NormalTok{(C)[}\DecValTok{1}\NormalTok{]}
  
  \ControlFlowTok{if}\NormalTok{(dim\_out }\OperatorTok{==}\StringTok{ }\DecValTok{1}\NormalTok{)\{}
\NormalTok{    all.Theta \textless{}{-}}\StringTok{ }\KeywordTok{array}\NormalTok{(}\KeywordTok{c}\NormalTok{(}\DecValTok{1}\OperatorTok{/}\NormalTok{Theta, Theta), }\KeywordTok{c}\NormalTok{(}\DecValTok{1}\NormalTok{,}\DecValTok{1}\NormalTok{,}\DecValTok{2}\NormalTok{))}
\NormalTok{    all.C \textless{}{-}}\StringTok{ }\KeywordTok{array}\NormalTok{(}\KeywordTok{c}\NormalTok{(C}\OperatorTok{*}\NormalTok{Theta, C), }\KeywordTok{c}\NormalTok{(}\DecValTok{1}\NormalTok{,}\DecValTok{1}\NormalTok{,}\DecValTok{2}\NormalTok{))}
\NormalTok{  \} }\ControlFlowTok{else}\NormalTok{ \{}
    
    \CommentTok{\# Create a matrix with 2\^{}(number of roots) rows...}
\NormalTok{    combi \textless{}{-}}\StringTok{ }\KeywordTok{matrix}\NormalTok{(}\DecValTok{0}\NormalTok{, }\DecValTok{2}\OperatorTok{\^{}}\NormalTok{dim\_out, dim\_out)}
    
\NormalTok{    basic \textless{}{-}}\StringTok{ }\KeywordTok{matrix}\NormalTok{(}\KeywordTok{c}\NormalTok{(}\DecValTok{0}\NormalTok{,}\DecValTok{1}\NormalTok{), }\DataTypeTok{ncol =} \DecValTok{1}\NormalTok{)}
    
    \CommentTok{\# Fill columns of possibly ridiculously large matrix}
    \ControlFlowTok{for}\NormalTok{(ix\_col }\ControlFlowTok{in} \DecValTok{1}\OperatorTok{:}\NormalTok{dim\_out)\{}
\NormalTok{      combi[,ix\_col] \textless{}{-}}\StringTok{ }\KeywordTok{rep}\NormalTok{(basic,}\DecValTok{2}\OperatorTok{\^{}}\NormalTok{(ix\_col}\DecValTok{{-}1}\NormalTok{)) }\OperatorTok{\%x\%}\StringTok{ }\KeywordTok{rep}\NormalTok{(}\DecValTok{1}\NormalTok{,}\DecValTok{2}\OperatorTok{\^{}}\NormalTok{(dim\_out}\OperatorTok{{-}}\NormalTok{ix\_col))}
\NormalTok{    \}}
    
    \CommentTok{\# Allocate all possible choices of zeros inside and outside the unit circle}
\NormalTok{    all.Theta \textless{}{-}}\StringTok{ }\KeywordTok{array}\NormalTok{(}\OtherTok{NaN}\NormalTok{, }\KeywordTok{c}\NormalTok{(dim\_out, dim\_out, }\DecValTok{2}\OperatorTok{\^{}}\NormalTok{dim\_out))}
\NormalTok{    all.C \textless{}{-}}\StringTok{ }\KeywordTok{array}\NormalTok{(}\OtherTok{NaN}\NormalTok{, }\KeywordTok{c}\NormalTok{(dim\_out, dim\_out, }\DecValTok{2}\OperatorTok{\^{}}\NormalTok{dim\_out))}
    
    \CommentTok{\# Compute all possible MA polynomials}
    \ControlFlowTok{for}\NormalTok{(ix\_flip\_choice }\ControlFlowTok{in} \DecValTok{1}\OperatorTok{:}\NormalTok{(}\DecValTok{2}\OperatorTok{\^{}}\NormalTok{dim\_out))\{}
\NormalTok{      Theta\_C\_tmp \textless{}{-}}\StringTok{ }\KeywordTok{compute.other.basic.form}\NormalTok{(Theta, C, combi[ix\_flip\_choice, ])}
\NormalTok{      all.Theta[, , ix\_flip\_choice] \textless{}{-}}\StringTok{ }\NormalTok{Theta\_C\_tmp}\OperatorTok{$}\NormalTok{Theta}
\NormalTok{      all.C[, , ix\_flip\_choice] \textless{}{-}}\StringTok{ }\NormalTok{Theta\_C\_tmp}\OperatorTok{$}\NormalTok{C}
\NormalTok{    \}}
\NormalTok{  \}}
  
  \KeywordTok{return}\NormalTok{(}\KeywordTok{list}\NormalTok{(}\DataTypeTok{all.Theta =}\NormalTok{ all.Theta,}
              \DataTypeTok{all.C =}\NormalTok{ all.C))}
\NormalTok{\}}
  
\NormalTok{\} }\ControlFlowTok{else}\NormalTok{ \{}
\CommentTok{\# This is the GMR version with original variable names and no added comments by BF}

\NormalTok{build.othonormal.basis \textless{}{-}}\StringTok{ }\ControlFlowTok{function}\NormalTok{(z)\{}
  \CommentTok{\# z is a vector of dimension n}
  \CommentTok{\# This function creates an orthonromal matrix K whose first column }
  \CommentTok{\# is the normalised value of z.}
\NormalTok{  n \textless{}{-}}\StringTok{ }\KeywordTok{length}\NormalTok{(z)}
\NormalTok{  I \textless{}{-}}\StringTok{ }\KeywordTok{diag}\NormalTok{(n)}
  \ControlFlowTok{if}\NormalTok{(z[}\DecValTok{1}\NormalTok{]}\OperatorTok{!=}\DecValTok{0}\NormalTok{)\{}
\NormalTok{    Z \textless{}{-}}\StringTok{ }\KeywordTok{matrix}\NormalTok{(I[,}\DecValTok{2}\OperatorTok{:}\NormalTok{n],n,n}\DecValTok{{-}1}\NormalTok{)}
\NormalTok{  \}}\ControlFlowTok{else}\NormalTok{\{}
\NormalTok{    Z \textless{}{-}}\StringTok{ }\KeywordTok{matrix}\NormalTok{(I[,}\DecValTok{1}\OperatorTok{:}\NormalTok{(n}\DecValTok{{-}1}\NormalTok{)],n,n}\DecValTok{{-}1}\NormalTok{)}
\NormalTok{  \}}
  \ControlFlowTok{if}\NormalTok{(}\KeywordTok{sum}\NormalTok{(z}\OperatorTok{\^{}}\DecValTok{2}\NormalTok{)}\OperatorTok{==}\DecValTok{0}\NormalTok{)\{}
\NormalTok{    K \textless{}{-}}\StringTok{ }\OtherTok{NaN}
    \KeywordTok{print}\NormalTok{(}\StringTok{"z should not be 0"}\NormalTok{)}
\NormalTok{  \}}\ControlFlowTok{else}\NormalTok{\{}
\NormalTok{    K \textless{}{-}}\StringTok{ }\KeywordTok{matrix}\NormalTok{(z}\OperatorTok{/}\KeywordTok{sqrt}\NormalTok{(}\KeywordTok{sum}\NormalTok{(z}\OperatorTok{\^{}}\DecValTok{2}\NormalTok{)),}\DataTypeTok{ncol=}\DecValTok{1}\NormalTok{)}
    \ControlFlowTok{for}\NormalTok{(i }\ControlFlowTok{in} \DecValTok{2}\OperatorTok{:}\NormalTok{n)\{}
\NormalTok{      y \textless{}{-}}\StringTok{ }\KeywordTok{matrix}\NormalTok{(Z[,i}\DecValTok{{-}1}\NormalTok{],}\DataTypeTok{ncol=}\DecValTok{1}\NormalTok{)}
\NormalTok{      X \textless{}{-}}\StringTok{ }\NormalTok{K[,}\DecValTok{1}\OperatorTok{:}\NormalTok{(i}\DecValTok{{-}1}\NormalTok{)]}
\NormalTok{      k \textless{}{-}}\StringTok{ }\NormalTok{(I }\OperatorTok{{-}}\StringTok{ }\NormalTok{X }\OperatorTok{\%*\%}\StringTok{ }\KeywordTok{solve}\NormalTok{(}\KeywordTok{t}\NormalTok{(X) }\OperatorTok{\%*\%}\StringTok{ }\NormalTok{X) }\OperatorTok{\%*\%}\StringTok{ }\KeywordTok{t}\NormalTok{(X)) }\OperatorTok{\%*\%}\StringTok{ }\NormalTok{y}
\NormalTok{      k \textless{}{-}}\StringTok{ }\NormalTok{k}\OperatorTok{/}\KeywordTok{sqrt}\NormalTok{(}\KeywordTok{sum}\NormalTok{(k}\OperatorTok{\^{}}\DecValTok{2}\NormalTok{))}
\NormalTok{      K \textless{}{-}}\StringTok{ }\KeywordTok{cbind}\NormalTok{(K,k)}
\NormalTok{    \}}
\NormalTok{  \}}
  \KeywordTok{return}\NormalTok{(K)}
\NormalTok{\}}

\NormalTok{compute.other.basic.form \textless{}{-}}\StringTok{ }\ControlFlowTok{function}\NormalTok{(Theta,C,w)\{}
  \CommentTok{\# w is a vector whose entries are 1 or {-}1 (see Lippi and Reichlin 1994)}
  \CommentTok{\# w is of dimension n, where Theta is n x n.}
  \CommentTok{\# If w\_i={-}1 then the output MA represention corresponds to an MA representation }
  \CommentTok{\# with the same spetral density as the input one, }
  \CommentTok{\# but the i\^{}th pole of det(I {-} Theta.L) would have been replaced }
  \CommentTok{\# by its inverse conjugate.}
  
\NormalTok{  n \textless{}{-}}\StringTok{ }\KeywordTok{dim}\NormalTok{(C)[}\DecValTok{1}\NormalTok{]}
  
\NormalTok{  poles.ini \textless{}{-}}\StringTok{ }\DecValTok{1}\OperatorTok{/}\KeywordTok{eigen}\NormalTok{(Theta)}\OperatorTok{$}\NormalTok{values}
  
\NormalTok{  Theta.k \textless{}{-}}\StringTok{ }\NormalTok{Theta}
\NormalTok{  C.k \textless{}{-}}\StringTok{ }\NormalTok{C}
  
  \ControlFlowTok{for}\NormalTok{(nb.pole }\ControlFlowTok{in} \DecValTok{1}\OperatorTok{:}\NormalTok{n)\{}
    \ControlFlowTok{if}\NormalTok{(w[nb.pole]}\OperatorTok{!=}\DecValTok{1}\NormalTok{)\{}
\NormalTok{      pole \textless{}{-}}\StringTok{ }\NormalTok{poles.ini[nb.pole]}
\NormalTok{      poles.k \textless{}{-}}\StringTok{ }\DecValTok{1}\OperatorTok{/}\KeywordTok{eigen}\NormalTok{(Theta.k)}\OperatorTok{$}\NormalTok{values}
\NormalTok{      indic.pole.k \textless{}{-}}\StringTok{ }\KeywordTok{which}\NormalTok{(}\KeywordTok{abs}\NormalTok{(poles.k}\OperatorTok{{-}}\NormalTok{pole)}\OperatorTok{==}\KeywordTok{min}\NormalTok{(}\KeywordTok{abs}\NormalTok{(poles.k}\OperatorTok{{-}}\NormalTok{pole)))[}\DecValTok{1}\NormalTok{]}
      
\NormalTok{      k \textless{}{-}}\StringTok{ }\KeywordTok{eigen}\NormalTok{(Theta.k)}\OperatorTok{$}\NormalTok{vectors[,indic.pole.k]}
\NormalTok{      k.star \textless{}{-}}\StringTok{ }\KeywordTok{solve}\NormalTok{(C.k) }\OperatorTok{\%*\%}\StringTok{ }\NormalTok{k}
      
\NormalTok{      K \textless{}{-}}\StringTok{ }\KeywordTok{build.othonormal.basis}\NormalTok{(k.star)}
      
\NormalTok{      CK.mod \textless{}{-}}\StringTok{ }\NormalTok{C.k }\OperatorTok{\%*\%}\StringTok{ }\NormalTok{K}
\NormalTok{      a \textless{}{-}}\StringTok{ }\NormalTok{CK.mod[,}\DecValTok{1}\NormalTok{]}
\NormalTok{      CK.mod[,}\DecValTok{1}\NormalTok{] \textless{}{-}}\StringTok{ }\OperatorTok{{-}}\StringTok{ }\NormalTok{a }\OperatorTok{*}\StringTok{ }\NormalTok{(}\DecValTok{1}\OperatorTok{/}\KeywordTok{Conj}\NormalTok{(pole))}
      
\NormalTok{      Theta.mod \textless{}{-}}\StringTok{ }\NormalTok{Theta.k }\OperatorTok{\%*\%}\StringTok{ }\NormalTok{C.k }\OperatorTok{\%*\%}\StringTok{ }\NormalTok{K}
\NormalTok{      Theta.mod[,}\DecValTok{1}\NormalTok{] \textless{}{-}}\StringTok{ }\OperatorTok{{-}}\StringTok{ }\NormalTok{a}
      
\NormalTok{      C.k \textless{}{-}}\StringTok{ }\NormalTok{CK.mod}
\NormalTok{      Theta.k \textless{}{-}}\StringTok{ }\NormalTok{Theta.mod }\OperatorTok{\%*\%}\StringTok{ }\KeywordTok{solve}\NormalTok{(CK.mod)}
\NormalTok{    \}}
\NormalTok{  \}}
  \KeywordTok{return}\NormalTok{(}\KeywordTok{list}\NormalTok{(}\DataTypeTok{Theta=}\NormalTok{Theta.k,}\DataTypeTok{C=}\NormalTok{C.k))}
\NormalTok{\}}

\NormalTok{compute.all.forms \textless{}{-}}\StringTok{ }\ControlFlowTok{function}\NormalTok{(Theta,C)\{}
  
\NormalTok{  n \textless{}{-}}\StringTok{ }\KeywordTok{dim}\NormalTok{(C)[}\DecValTok{1}\NormalTok{]}
  
  \ControlFlowTok{if}\NormalTok{(n}\OperatorTok{==}\DecValTok{1}\NormalTok{)\{}
\NormalTok{    all.Theta \textless{}{-}}\StringTok{ }\KeywordTok{array}\NormalTok{(}\KeywordTok{c}\NormalTok{(}\DecValTok{1}\OperatorTok{/}\NormalTok{Theta,Theta),}\KeywordTok{c}\NormalTok{(}\DecValTok{1}\NormalTok{,}\DecValTok{1}\NormalTok{,}\DecValTok{2}\NormalTok{))}
\NormalTok{    all.C \textless{}{-}}\StringTok{ }\KeywordTok{array}\NormalTok{(}\KeywordTok{c}\NormalTok{(C}\OperatorTok{*}\NormalTok{Theta,C),}\KeywordTok{c}\NormalTok{(}\DecValTok{1}\NormalTok{,}\DecValTok{1}\NormalTok{,}\DecValTok{2}\NormalTok{))}
\NormalTok{  \}}\ControlFlowTok{else}\NormalTok{\{}
\NormalTok{    combi \textless{}{-}}\StringTok{ }\KeywordTok{matrix}\NormalTok{(}\DecValTok{0}\NormalTok{,}\DecValTok{2}\OperatorTok{\^{}}\NormalTok{n,n)}
\NormalTok{    basic \textless{}{-}}\StringTok{ }\KeywordTok{matrix}\NormalTok{(}\KeywordTok{c}\NormalTok{(}\DecValTok{0}\NormalTok{,}\DecValTok{1}\NormalTok{),}\DataTypeTok{ncol=}\DecValTok{1}\NormalTok{)}
    \ControlFlowTok{for}\NormalTok{(i }\ControlFlowTok{in} \DecValTok{1}\OperatorTok{:}\NormalTok{n)\{}
\NormalTok{      combi[,i] \textless{}{-}}\StringTok{ }\KeywordTok{rep}\NormalTok{(basic,}\DecValTok{2}\OperatorTok{\^{}}\NormalTok{(i}\DecValTok{{-}1}\NormalTok{)) }\OperatorTok{\%x\%}\StringTok{ }\KeywordTok{rep}\NormalTok{(}\DecValTok{1}\NormalTok{,}\DecValTok{2}\OperatorTok{\^{}}\NormalTok{(n}\OperatorTok{{-}}\NormalTok{i))}
\NormalTok{    \}}
    
\NormalTok{    all.Theta \textless{}{-}}\StringTok{ }\KeywordTok{array}\NormalTok{(}\OtherTok{NaN}\NormalTok{,}\KeywordTok{c}\NormalTok{(n,n,}\DecValTok{2}\OperatorTok{\^{}}\NormalTok{n))}
\NormalTok{    all.C \textless{}{-}}\StringTok{ }\KeywordTok{array}\NormalTok{(}\OtherTok{NaN}\NormalTok{,}\KeywordTok{c}\NormalTok{(n,n,}\DecValTok{2}\OperatorTok{\^{}}\NormalTok{n))}
    
    \ControlFlowTok{for}\NormalTok{(i }\ControlFlowTok{in} \DecValTok{1}\OperatorTok{:}\NormalTok{(}\DecValTok{2}\OperatorTok{\^{}}\NormalTok{n))\{}
\NormalTok{      aux \textless{}{-}}\StringTok{ }\KeywordTok{compute.other.basic.form}\NormalTok{(Theta,C,combi[i,])}
\NormalTok{      all.Theta[,,i] \textless{}{-}}\StringTok{ }\NormalTok{aux}\OperatorTok{$}\NormalTok{Theta}
\NormalTok{      all.C[,,i] \textless{}{-}}\StringTok{ }\NormalTok{aux}\OperatorTok{$}\NormalTok{C}
\NormalTok{    \}}
\NormalTok{  \}}
  
  \KeywordTok{return}\NormalTok{(}\KeywordTok{list}\NormalTok{(}\DataTypeTok{all.Theta=}\NormalTok{all.Theta,}\DataTypeTok{all.C=}\NormalTok{all.C))}
\NormalTok{\}}
  
\NormalTok{\}}
\end{Highlighting}
\end{Shaded}

\hypertarget{code-for-examples}{%
\section{Code for Examples}\label{code-for-examples}}

Using GMR's implementation of Blaschke matrices, we provide examples for
the cases described in the main text.

\hypertarget{purely-complex-case-1}{%
\subsection{Purely Complex Case}\label{purely-complex-case-1}}

We define a particular matrix of the kind
\(\left(\begin{smallmatrix} 0 & b \\ -b & 0 \end{smallmatrix}\right)\).

\begin{Shaded}
\begin{Highlighting}[]
\ControlFlowTok{if}\NormalTok{(params}\OperatorTok{$}\NormalTok{generate\_random\_examples)\{}
  \CommentTok{\# Randomly generate imaginary part:}
\NormalTok{  a =}\StringTok{ }\DecValTok{0}
\NormalTok{  b =}\StringTok{ }\KeywordTok{floor}\NormalTok{(}\KeywordTok{runif}\NormalTok{(}\DecValTok{1}\NormalTok{, }\DataTypeTok{min =} \DecValTok{{-}1}\NormalTok{, }\DataTypeTok{max =} \DecValTok{1}\NormalTok{)}\OperatorTok{*}\DecValTok{10}\OperatorTok{+}\DecValTok{1}\NormalTok{)}
\NormalTok{\} }\ControlFlowTok{else}\NormalTok{ \{}
\NormalTok{  a =}\StringTok{ }\DecValTok{0}
\NormalTok{  b =}\StringTok{ }\DecValTok{1}
\NormalTok{\}}

\CommentTok{\# MA(1) coefficient matrix: }
\NormalTok{(}\DataTypeTok{Theta =} \KeywordTok{matrix}\NormalTok{(}\KeywordTok{c}\NormalTok{(a,b,}\OperatorTok{{-}}\NormalTok{b,a), }\DecValTok{2}\NormalTok{, }\DecValTok{2}\NormalTok{))}
\CommentTok{\#\#      [,1] [,2]}
\CommentTok{\#\# [1,]    0   {-}1}
\CommentTok{\#\# [2,]    1    0}

\CommentTok{\# Static shock transmission matrix: }
\NormalTok{(}\DataTypeTok{C =} \KeywordTok{diag}\NormalTok{(}\DecValTok{2}\NormalTok{))}
\CommentTok{\#\#      [,1] [,2]}
\CommentTok{\#\# [1,]    1    0}
\CommentTok{\#\# [2,]    0    1}
\end{Highlighting}
\end{Shaded}

Next, we calculate its eigenvalues.

\begin{Shaded}
\begin{Highlighting}[]
\CommentTok{\# Its eigenvalues: }
\NormalTok{(}\DataTypeTok{ev\_Theta =} \KeywordTok{eigen}\NormalTok{(Theta)}\OperatorTok{$}\NormalTok{values)}
\CommentTok{\#\# [1] 0+1i 0{-}1i}

\CommentTok{\# Determinantal roots of the MA polynomial: }
\NormalTok{(}\DataTypeTok{ev\_Theta\_inv =}\NormalTok{ ev\_Theta}\OperatorTok{\^{}}\NormalTok{(}\OperatorTok{{-}}\DecValTok{1}\NormalTok{))}
\CommentTok{\#\# [1] 0{-}1i 0+1i}
\end{Highlighting}
\end{Shaded}

Calling GMR's procedure results in an error.

\begin{Shaded}
\begin{Highlighting}[]
\NormalTok{(}\DataTypeTok{ex\_trivial\_rand\_purelycplx\_conj\_10 =} \KeywordTok{try}\NormalTok{(}\KeywordTok{compute.other.basic.form}\NormalTok{(}\DataTypeTok{Theta =}\NormalTok{ Theta, }\DataTypeTok{C =}\NormalTok{ C, }
                                                                   \DataTypeTok{w =} \KeywordTok{c}\NormalTok{(}\DecValTok{1}\NormalTok{,}\DecValTok{0}\NormalTok{))))}
\CommentTok{\#\# [1] "z should not be 0"}
\CommentTok{\#\# Error in C\_step \%*\% trafo\_orth : non{-}conformable arguments}
\CommentTok{\#\# [1] "Error in C\_step \%*\% trafo\_orth : non{-}conformable arguments\textbackslash{}n"}
\CommentTok{\#\# attr(,"class")}
\CommentTok{\#\# [1] "try{-}error"}
\CommentTok{\#\# attr(,"condition")}
\CommentTok{\#\# \textless{}simpleError in C\_step \%*\% trafo\_orth: non{-}conformable arguments\textgreater{}}
\end{Highlighting}
\end{Shaded}

\hypertarget{skew-symmetric-complex-case-1}{%
\subsection{Skew-Symmetric Complex
Case}\label{skew-symmetric-complex-case-1}}

Next, we generate two matrices of the kind
\(\left(\begin{smallmatrix} a & b \\ -b & a \end{smallmatrix}\right)\).
For \(a=b=1\) an error is thrown, whereas for \(a=2\) and \(b=-6\)
obviously incorrect results are obtained.

\begin{Shaded}
\begin{Highlighting}[]
\CommentTok{\# Choose parameter values such that an error is thrown}
\NormalTok{a =}\StringTok{ }\DecValTok{1}
\NormalTok{b =}\StringTok{ }\DecValTok{1}

\CommentTok{\# MA(1) coefficient matrix:}
\NormalTok{(}\DataTypeTok{Theta =} \KeywordTok{matrix}\NormalTok{(}\KeywordTok{c}\NormalTok{(a,b,}\OperatorTok{{-}}\NormalTok{b,a), }\DecValTok{2}\NormalTok{, }\DecValTok{2}\NormalTok{))}
\CommentTok{\#\#      [,1] [,2]}
\CommentTok{\#\# [1,]    1   {-}1}
\CommentTok{\#\# [2,]    1    1}

\CommentTok{\# Static shock transmission matrix: }
\NormalTok{(}\DataTypeTok{C =} \KeywordTok{diag}\NormalTok{(}\DecValTok{2}\NormalTok{))}
\CommentTok{\#\#      [,1] [,2]}
\CommentTok{\#\# [1,]    1    0}
\CommentTok{\#\# [2,]    0    1}
\end{Highlighting}
\end{Shaded}

Calling GMR's procedure results in an error.

\begin{Shaded}
\begin{Highlighting}[]
\NormalTok{(}\DataTypeTok{ex1\_skewsymm\_10 =} \KeywordTok{try}\NormalTok{(}\KeywordTok{compute.other.basic.form}\NormalTok{(}\DataTypeTok{Theta =}\NormalTok{ Theta, }\DataTypeTok{C =}\NormalTok{ C, }
                                                \DataTypeTok{w =} \KeywordTok{c}\NormalTok{(}\DecValTok{1}\NormalTok{,}\DecValTok{0}\NormalTok{))))}
\CommentTok{\#\# [1] "z should not be 0"}
\CommentTok{\#\# Error in C\_step \%*\% trafo\_orth : non{-}conformable arguments}
\CommentTok{\#\# [1] "Error in C\_step \%*\% trafo\_orth : non{-}conformable arguments\textbackslash{}n"}
\CommentTok{\#\# attr(,"class")}
\CommentTok{\#\# [1] "try{-}error"}
\CommentTok{\#\# attr(,"condition")}
\CommentTok{\#\# \textless{}simpleError in C\_step \%*\% trafo\_orth: non{-}conformable arguments\textgreater{}}
\end{Highlighting}
\end{Shaded}

While the next example does not throw an error, the obtained matrices
are not all-pass and the eigenvalues are not mirrored at the unit
circle.

\begin{Shaded}
\begin{Highlighting}[]
\CommentTok{\# Choose parameter values such that no error is thrown but that results are incorrect}
\NormalTok{a =}\StringTok{ }\DecValTok{2}
\NormalTok{b =}\StringTok{ }\DecValTok{{-}6}

\CommentTok{\# MA(1) coefficient matrix:}
\NormalTok{(}\DataTypeTok{Theta =} \KeywordTok{matrix}\NormalTok{(}\KeywordTok{c}\NormalTok{(a,b,}\OperatorTok{{-}}\NormalTok{b,a), }\DecValTok{2}\NormalTok{, }\DecValTok{2}\NormalTok{))}
\CommentTok{\#\#      [,1] [,2]}
\CommentTok{\#\# [1,]    2    6}
\CommentTok{\#\# [2,]   {-}6    2}

\CommentTok{\# Static shock transmission matrix: }
\NormalTok{(}\DataTypeTok{C =} \KeywordTok{diag}\NormalTok{(}\DecValTok{2}\NormalTok{))}
\CommentTok{\#\#      [,1] [,2]}
\CommentTok{\#\# [1,]    1    0}
\CommentTok{\#\# [2,]    0    1}
\end{Highlighting}
\end{Shaded}

Calling GMR's procedure results in obviously incorrect values.

\begin{Shaded}
\begin{Highlighting}[]
\NormalTok{(}\DataTypeTok{ex2\_skewsymm\_10 =} \KeywordTok{try}\NormalTok{(}\KeywordTok{compute.other.basic.form}\NormalTok{(}\DataTypeTok{Theta =}\NormalTok{ Theta, }\DataTypeTok{C =}\NormalTok{ C, }
                                                \DataTypeTok{w =} \KeywordTok{c}\NormalTok{(}\DecValTok{1}\NormalTok{,}\DecValTok{0}\NormalTok{))))}
\CommentTok{\#\# $Theta}
\CommentTok{\#\#                             [,1]                       [,2]}
\CommentTok{\#\# [1,] {-}2.639553e+15+1.273073e+16i 1.273073e+16+2.639553e+15i}
\CommentTok{\#\# [2,]  1.273073e+16+2.639553e+15i 2.639553e+15{-}1.273073e+16i}
\CommentTok{\#\# }
\CommentTok{\#\# $C}
\CommentTok{\#\#                       [,1]               [,2]}
\CommentTok{\#\# [1,]  {-}94906266{-}284718797i        0+47453133i}
\CommentTok{\#\# [2,] {-}284718797+ 94906266i 47453133+       0i}
\end{Highlighting}
\end{Shaded}

There is no apparent connection of the eigenvalues of the original MA(1)
coefficient to the ones of the newly transformed system.

\begin{Shaded}
\begin{Highlighting}[]
\CommentTok{\# Eigenvalues of original MA(1) coefficient matrix }
\KeywordTok{eigen}\NormalTok{(Theta)}\OperatorTok{$}\NormalTok{values}
\CommentTok{\#\# [1] 2+6i 2{-}6i}

\CommentTok{\# Eigenvalues of original MA(1) coefficient matrix }
\KeywordTok{eigen}\NormalTok{(ex2\_skewsymm\_}\DecValTok{10}\OperatorTok{$}\NormalTok{Theta)}\OperatorTok{$}\NormalTok{values}
\CommentTok{\#\# [1] {-}295256766{-}93315703i  295256765+93315703i}
\end{Highlighting}
\end{Shaded}

\hypertarget{general-complex-case-1}{%
\subsection{General Complex Case}\label{general-complex-case-1}}

Here, we consider matrices of the kind
\(\left(\begin{smallmatrix} a & b + c \\ -b & a \end{smallmatrix}\right)\),
where \(c \cdot b > 0\) such that we obtain complex-valued eigenvalues.

\begin{Shaded}
\begin{Highlighting}[]
\ControlFlowTok{if}\NormalTok{(params}\OperatorTok{$}\NormalTok{generate\_random\_examples)\{}
  \CommentTok{\# Randomly generate imaginary part:}
\NormalTok{  a =}\StringTok{ }\KeywordTok{floor}\NormalTok{(}\KeywordTok{runif}\NormalTok{(}\DecValTok{1}\NormalTok{, }\DataTypeTok{min =} \DecValTok{{-}1}\NormalTok{, }\DataTypeTok{max =} \DecValTok{1}\NormalTok{)}\OperatorTok{*}\DecValTok{10}\OperatorTok{+}\DecValTok{1}\NormalTok{)}
\NormalTok{  b =}\StringTok{ }\KeywordTok{floor}\NormalTok{(}\KeywordTok{runif}\NormalTok{(}\DecValTok{1}\NormalTok{, }\DataTypeTok{min =} \DecValTok{{-}1}\NormalTok{, }\DataTypeTok{max =} \DecValTok{1}\NormalTok{)}\OperatorTok{*}\DecValTok{10}\OperatorTok{+}\DecValTok{1}\NormalTok{)}
  \ControlFlowTok{while}\NormalTok{(b }\OperatorTok{==}\StringTok{ }\DecValTok{0}\NormalTok{)\{}
\NormalTok{    b =}\StringTok{ }\KeywordTok{floor}\NormalTok{(}\KeywordTok{runif}\NormalTok{(}\DecValTok{1}\NormalTok{, }\DataTypeTok{min =} \DecValTok{{-}1}\NormalTok{, }\DataTypeTok{max =} \DecValTok{1}\NormalTok{)}\OperatorTok{*}\DecValTok{10}\OperatorTok{+}\DecValTok{1}\NormalTok{)}
\NormalTok{  \}}
\NormalTok{  c =}\StringTok{ }\KeywordTok{floor}\NormalTok{(}\KeywordTok{runif}\NormalTok{(}\DecValTok{1}\NormalTok{, }\DataTypeTok{min =} \DecValTok{{-}1}\NormalTok{, }\DataTypeTok{max =} \DecValTok{1}\NormalTok{)}\OperatorTok{*}\DecValTok{10}\OperatorTok{+}\DecValTok{1}\NormalTok{)}
    \ControlFlowTok{while}\NormalTok{(c }\OperatorTok{==}\StringTok{ }\DecValTok{0}\NormalTok{)\{}
\NormalTok{    c =}\StringTok{ }\KeywordTok{floor}\NormalTok{(}\KeywordTok{runif}\NormalTok{(}\DecValTok{1}\NormalTok{, }\DataTypeTok{min =} \DecValTok{{-}1}\NormalTok{, }\DataTypeTok{max =} \DecValTok{1}\NormalTok{)}\OperatorTok{*}\DecValTok{10}\OperatorTok{+}\DecValTok{1}\NormalTok{)}
\NormalTok{    \}}
  \ControlFlowTok{if}\NormalTok{(b}\OperatorTok{*}\NormalTok{c }\OperatorTok{\textless{}}\StringTok{ }\DecValTok{0}\NormalTok{)\{}
\NormalTok{    c =}\StringTok{ }\OperatorTok{{-}}\NormalTok{c}
\NormalTok{  \}}
\NormalTok{\} }\ControlFlowTok{else}\NormalTok{ \{}
\NormalTok{  a =}\StringTok{ }\DecValTok{4}
\NormalTok{  b =}\StringTok{ }\DecValTok{3}
\NormalTok{  c =}\StringTok{ }\DecValTok{2}
\NormalTok{\}}

\CommentTok{\# MA(1) coefficient matrix: }
\NormalTok{(}\DataTypeTok{Theta =} \KeywordTok{matrix}\NormalTok{(}\KeywordTok{c}\NormalTok{(a, b}\OperatorTok{+}\NormalTok{c, }\OperatorTok{{-}}\NormalTok{b, a), }\DecValTok{2}\NormalTok{, }\DecValTok{2}\NormalTok{))}
\CommentTok{\#\#      [,1] [,2]}
\CommentTok{\#\# [1,]    4   {-}3}
\CommentTok{\#\# [2,]    5    4}

\CommentTok{\# Eigenvalues of Theta:}
\KeywordTok{eigen}\NormalTok{(Theta)}\OperatorTok{$}\NormalTok{values}
\CommentTok{\#\# [1] 4+3.872983i 4{-}3.872983i}

\CommentTok{\# Static shock transmission matrix: }
\NormalTok{(}\DataTypeTok{C =} \KeywordTok{diag}\NormalTok{(}\DecValTok{2}\NormalTok{))}
\CommentTok{\#\#      [,1] [,2]}
\CommentTok{\#\# [1,]    1    0}
\CommentTok{\#\# [2,]    0    1}
\end{Highlighting}
\end{Shaded}

While no error is thrown, we will show that when both complex-conjugated
roots are mirrored at the unit circle, the result does not constitute an
observationally equivalent matrix polynomial (even though the
eigenvalues of the original MA(1) coefficient matrix are mirrored
correctly). First, we mirror both eigenvalues.

\begin{Shaded}
\begin{Highlighting}[]
\NormalTok{(}\DataTypeTok{ex\_nonskew\_00 =} \KeywordTok{try}\NormalTok{(}\KeywordTok{compute.other.basic.form}\NormalTok{(}\DataTypeTok{Theta =}\NormalTok{ Theta, }\DataTypeTok{C =}\NormalTok{ C, }
                                              \DataTypeTok{w =} \KeywordTok{c}\NormalTok{(}\DecValTok{0}\NormalTok{,}\DecValTok{0}\NormalTok{))))}
\CommentTok{\#\# $Theta}
\CommentTok{\#\#                [,1]         [,2]}
\CommentTok{\#\# [1,]  0.05990783+0i 0.1105991+0i}
\CommentTok{\#\# [2,] {-}0.18433180+0i 0.1981567{-}0i}
\CommentTok{\#\# }
\CommentTok{\#\# $C}
\CommentTok{\#\#                     [,1]                [,2]}
\CommentTok{\#\# [1,]  3.600595{-}6.338657i {-}4.909903{-}4.648348i}
\CommentTok{\#\# [2,] {-}4.909903{-}8.028965i {-}6.219210+6.338657i}
\end{Highlighting}
\end{Shaded}

The eigenvalues of the transformed MA(1) coefficient matrix are the
inverses of the eigenvalues of the original MA(1) coefficient matrix.

\begin{Shaded}
\begin{Highlighting}[]
\CommentTok{\# Eigenvalues of original MA(1) coefficient matrix }
\KeywordTok{eigen}\NormalTok{(Theta)}\OperatorTok{$}\NormalTok{values}
\CommentTok{\#\# [1] 4+3.872983i 4{-}3.872983i}

\CommentTok{\# Inverses of eigenvalues of original MA(1) coefficient matrix }
\KeywordTok{eigen}\NormalTok{(Theta)}\OperatorTok{$}\NormalTok{values}\OperatorTok{\^{}}\NormalTok{(}\OperatorTok{{-}}\DecValTok{1}\NormalTok{)}
\CommentTok{\#\# [1] 0.1290323{-}0.1249349i 0.1290323+0.1249349i}

\CommentTok{\# Eigenvalues of transformed MA(1) coefficient matrix }
\KeywordTok{eigen}\NormalTok{(ex\_nonskew\_}\DecValTok{00}\OperatorTok{$}\NormalTok{Theta)}\OperatorTok{$}\NormalTok{values}
\CommentTok{\#\# [1] 0.1290323{-}0.1249349i 0.1290323+0.1249349i}
\end{Highlighting}
\end{Shaded}

However, the original and transformed MA(1) polynomial matrices are not
observationally equivalent. This is a consequence of the fact that the
complex nature of the eigenvectors is ignored in GMR's function
\emph{build.orthonormal.basis(\(\ldots\))}. In order to show this, we
evaluate both matrix polynomials on the unit circle.

\begin{Shaded}
\begin{Highlighting}[]
\CommentTok{\# Generate (equally{-}spaced) points on the unit circle:}
\NormalTok{z =}\StringTok{ }\KeywordTok{exp}\NormalTok{(1i}\OperatorTok{*}\DecValTok{2}\OperatorTok{*}\NormalTok{pi}\OperatorTok{*}\KeywordTok{seq}\NormalTok{(}\DecValTok{0}\NormalTok{, }\DecValTok{1}\NormalTok{, }\DataTypeTok{length.out =} \DecValTok{11}\NormalTok{))[}\OperatorTok{{-}}\DecValTok{11}\NormalTok{]}

\CommentTok{\# Functions to evaluate original and transformed MA(1) polynomial matrices:}
\NormalTok{polymat\_original =}\StringTok{ }\ControlFlowTok{function}\NormalTok{(z)\{}
\NormalTok{  (}\KeywordTok{diag}\NormalTok{(}\DecValTok{2}\NormalTok{) }\OperatorTok{{-}}\StringTok{ }\NormalTok{Theta }\OperatorTok{*}\StringTok{ }\NormalTok{z) }\OperatorTok{\%*\%}\StringTok{ }\NormalTok{C}
\NormalTok{\}}

\NormalTok{polymat\_transformed =}\StringTok{ }\ControlFlowTok{function}\NormalTok{(z)\{}
\NormalTok{  (}\KeywordTok{diag}\NormalTok{(}\DecValTok{2}\NormalTok{) }\OperatorTok{{-}}\StringTok{ }\NormalTok{ex\_nonskew\_}\DecValTok{00}\OperatorTok{$}\NormalTok{Theta }\OperatorTok{*}\StringTok{ }\NormalTok{z) }\OperatorTok{\%*\%}\StringTok{ }\NormalTok{ex\_nonskew\_}\DecValTok{00}\OperatorTok{$}\NormalTok{C}
\NormalTok{\}}

\CommentTok{\# Function for comparing polynomial matrices evaluated on the unit circle}
\CommentTok{\# in terms of many different distance measures}
\NormalTok{compare\_original\_vs\_transformed =}\StringTok{ }\ControlFlowTok{function}\NormalTok{(original, transformed)\{}
  
  \CommentTok{\# Since the spectral densities of the original and transformed polynomial matrix }
  \CommentTok{\# should be identical, }
  \CommentTok{\# we evaluated them on various points on the unit circle, }
  \CommentTok{\# calculate the value of the spectral density for the original and transformed one,}
  \CommentTok{\# and compare their values}
  
\NormalTok{  original\_herm =}\StringTok{ }\NormalTok{original }\OperatorTok{\%*\%}\StringTok{ }\KeywordTok{Conj}\NormalTok{(}\KeywordTok{t}\NormalTok{(original))}
\NormalTok{  transformed\_herm =}\StringTok{ }\NormalTok{transformed }\OperatorTok{\%*\%}\StringTok{ }\KeywordTok{Conj}\NormalTok{(}\KeywordTok{t}\NormalTok{(transformed))}
\NormalTok{  diff\_herm =}\StringTok{ }\NormalTok{original\_herm }\OperatorTok{{-}}\StringTok{ }\NormalTok{transformed\_herm}
  
  \KeywordTok{return}\NormalTok{(}\KeywordTok{list}\NormalTok{(}\DataTypeTok{original\_herm =}\NormalTok{ original\_herm,}
              \DataTypeTok{transformed\_herm =}\NormalTok{ transformed\_herm,}
              \DataTypeTok{diff\_herm=}\NormalTok{ diff\_herm,}
              \DataTypeTok{dist\_herm =} \KeywordTok{sum}\NormalTok{(diff\_herm }\OperatorTok{*}\StringTok{ }\KeywordTok{Conj}\NormalTok{(diff\_herm)))}
\NormalTok{         )}
\NormalTok{\}}

\ControlFlowTok{for}\NormalTok{ (ix\_z }\ControlFlowTok{in} \KeywordTok{seq\_along}\NormalTok{(z))\{}
\NormalTok{  original\_evaluated =}\StringTok{ }\KeywordTok{polymat\_original}\NormalTok{(z[ix\_z])}
\NormalTok{  transformed\_evaluated =}\StringTok{ }\KeywordTok{polymat\_transformed}\NormalTok{(z[ix\_z])}
  
\NormalTok{  comparison\_out\_list =}\StringTok{ }\KeywordTok{compare\_original\_vs\_transformed}\NormalTok{(original\_evaluated, transformed\_evaluated)}
  
  \KeywordTok{cat}\NormalTok{(}\KeywordTok{paste0}\NormalTok{(}\StringTok{"Distance measure when evaluated at exp(2i pi x) for x equal to "}\NormalTok{, (ix\_z }\OperatorTok{{-}}\StringTok{ }\DecValTok{1}\NormalTok{)}\OperatorTok{/}\DecValTok{10}\NormalTok{, }\StringTok{" is : "}\NormalTok{, comparison\_out\_list}\OperatorTok{$}\NormalTok{dist\_herm, }\StringTok{"}\CharTok{\textbackslash{}n}\StringTok{"}\NormalTok{))}
\NormalTok{\}}
\CommentTok{\#\# Distance measure when evaluated at exp(2i pi x) for x equal to 0 is : 29924.8979591836+0i}
\CommentTok{\#\# Distance measure when evaluated at exp(2i pi x) for x equal to 0.1 is : 19163.9224190856+0i}
\CommentTok{\#\# Distance measure when evaluated at exp(2i pi x) for x equal to 0.2 is : 19947.6366425084+0i}
\CommentTok{\#\# Distance measure when evaluated at exp(2i pi x) for x equal to 0.3 is : 32593.8082149725+0i}
\CommentTok{\#\# Distance measure when evaluated at exp(2i pi x) for x equal to 0.4 is : 58200.1756711532+0i}
\CommentTok{\#\# Distance measure when evaluated at exp(2i pi x) for x equal to 0.5 is : 90032.6530612241+0i}
\CommentTok{\#\# Distance measure when evaluated at exp(2i pi x) for x equal to 0.6 is : 111887.119096838+0i}
\CommentTok{\#\# Distance measure when evaluated at exp(2i pi x) for x equal to 0.7 is : 109869.165310613+0i}
\CommentTok{\#\# Distance measure when evaluated at exp(2i pi x) for x equal to 0.8 is : 85366.7012425608+0i}
\CommentTok{\#\# Distance measure when evaluated at exp(2i pi x) for x equal to 0.9 is : 53666.9816063492+0i}
\end{Highlighting}
\end{Shaded}

Even if these issues were solved, the matrices would still not be
real-valued and therefore outside of GMR's model class.

\hypertarget{effects-of-discarding-imaginary-part}{%
\subsection{Effects of Discarding Imaginary
Part}\label{effects-of-discarding-imaginary-part}}

Here, we show that the number of roots inside the unit circle changes
when the imaginary part of the MA parameter matrix is discarded.

\begin{Shaded}
\begin{Highlighting}[]
\ControlFlowTok{if}\NormalTok{(params}\OperatorTok{$}\NormalTok{generate\_random\_examples)\{}
  \CommentTok{\# Randomly generate imaginary part:}
\NormalTok{  a =}\StringTok{ }\KeywordTok{floor}\NormalTok{(}\KeywordTok{runif}\NormalTok{(}\DecValTok{1}\NormalTok{, }\DataTypeTok{min =} \DecValTok{{-}1}\NormalTok{, }\DataTypeTok{max =} \DecValTok{1}\NormalTok{)}\OperatorTok{*}\DecValTok{10}\OperatorTok{+}\DecValTok{1}\NormalTok{)}
\NormalTok{  b =}\StringTok{ }\KeywordTok{floor}\NormalTok{(}\KeywordTok{runif}\NormalTok{(}\DecValTok{1}\NormalTok{, }\DataTypeTok{min =} \DecValTok{{-}1}\NormalTok{, }\DataTypeTok{max =} \DecValTok{1}\NormalTok{)}\OperatorTok{*}\DecValTok{10}\OperatorTok{+}\DecValTok{1}\NormalTok{)}
  \ControlFlowTok{while}\NormalTok{(b }\OperatorTok{==}\StringTok{ }\DecValTok{0}\NormalTok{)\{}
\NormalTok{    b =}\StringTok{ }\KeywordTok{floor}\NormalTok{(}\KeywordTok{runif}\NormalTok{(}\DecValTok{1}\NormalTok{, }\DataTypeTok{min =} \DecValTok{{-}1}\NormalTok{, }\DataTypeTok{max =} \DecValTok{1}\NormalTok{)}\OperatorTok{*}\DecValTok{10}\OperatorTok{+}\DecValTok{1}\NormalTok{)}
\NormalTok{  \}}
\NormalTok{  c =}\StringTok{ }\KeywordTok{floor}\NormalTok{(}\KeywordTok{runif}\NormalTok{(}\DecValTok{1}\NormalTok{, }\DataTypeTok{min =} \DecValTok{{-}1}\NormalTok{, }\DataTypeTok{max =} \DecValTok{1}\NormalTok{)}\OperatorTok{*}\DecValTok{10}\OperatorTok{+}\DecValTok{1}\NormalTok{)}
    \ControlFlowTok{while}\NormalTok{(c }\OperatorTok{==}\StringTok{ }\DecValTok{0}\NormalTok{)\{}
\NormalTok{    c =}\StringTok{ }\KeywordTok{floor}\NormalTok{(}\KeywordTok{runif}\NormalTok{(}\DecValTok{1}\NormalTok{, }\DataTypeTok{min =} \DecValTok{{-}1}\NormalTok{, }\DataTypeTok{max =} \DecValTok{1}\NormalTok{)}\OperatorTok{*}\DecValTok{10}\OperatorTok{+}\DecValTok{1}\NormalTok{)}
\NormalTok{    \}}
  \ControlFlowTok{if}\NormalTok{(b}\OperatorTok{*}\NormalTok{c }\OperatorTok{\textless{}}\StringTok{ }\DecValTok{0}\NormalTok{)\{}
\NormalTok{    c =}\StringTok{ }\OperatorTok{{-}}\NormalTok{c}
\NormalTok{  \}}
\NormalTok{\} }\ControlFlowTok{else}\NormalTok{ \{}
\NormalTok{  a =}\StringTok{ }\DecValTok{2}
\NormalTok{  b =}\StringTok{ }\DecValTok{2}
\NormalTok{  c =}\StringTok{ }\DecValTok{4}
\NormalTok{\}}

\CommentTok{\# MA(1) coefficient matrix: }
\NormalTok{(}\DataTypeTok{Theta =} \KeywordTok{matrix}\NormalTok{(}\KeywordTok{c}\NormalTok{(a, b}\OperatorTok{+}\NormalTok{c, }\OperatorTok{{-}}\NormalTok{b, a), }\DecValTok{2}\NormalTok{, }\DecValTok{2}\NormalTok{))}
\CommentTok{\#\#      [,1] [,2]}
\CommentTok{\#\# [1,]    2   {-}2}
\CommentTok{\#\# [2,]    6    2}

\CommentTok{\# Eigenvalues of Theta:}
\KeywordTok{eigen}\NormalTok{(Theta)}\OperatorTok{$}\NormalTok{values}
\CommentTok{\#\# [1] 2+3.464102i 2{-}3.464102i}

\CommentTok{\# Static shock transmission matrix: }
\NormalTok{(}\DataTypeTok{C =} \KeywordTok{diag}\NormalTok{(}\DecValTok{2}\NormalTok{))}
\CommentTok{\#\#      [,1] [,2]}
\CommentTok{\#\# [1,]    1    0}
\CommentTok{\#\# [2,]    0    1}
\end{Highlighting}
\end{Shaded}

Mirroring one determinantal root inside the unit circle results in a
complex-valued parameter matrix. The complex-valued parameter matrix has
one determinantal root inside the unit circle. However, when the
imaginary parts are discarded, all roots are outside the unit circle.

\begin{Shaded}
\begin{Highlighting}[]
\NormalTok{(}\DataTypeTok{ex\_nonskew\_10 =} \KeywordTok{try}\NormalTok{(}\KeywordTok{compute.other.basic.form}\NormalTok{(}\DataTypeTok{Theta =}\NormalTok{ Theta, }\DataTypeTok{C =}\NormalTok{ C, }
                                              \DataTypeTok{w =} \KeywordTok{c}\NormalTok{(}\DecValTok{1}\NormalTok{,}\DecValTok{0}\NormalTok{))))}
\CommentTok{\#\# $Theta}
\CommentTok{\#\#                  [,1]              [,2]}
\CommentTok{\#\# [1,] 2.9375{-}1.623798i  0.8125+1.623798i}
\CommentTok{\#\# [2,] 8.8125+1.623798i {-}0.8125+4.871393i}
\CommentTok{\#\# }
\CommentTok{\#\# $C}
\CommentTok{\#\#                    [,1]                [,2]}
\CommentTok{\#\# [1,] {-}2.44949+1.414214i 1.224745+0.0000000i}
\CommentTok{\#\# [2,] {-}2.44949{-}4.242641i 0.000000+0.7071068i}
\KeywordTok{eigen}\NormalTok{(ex\_nonskew\_}\DecValTok{10}\OperatorTok{$}\NormalTok{Theta)}\OperatorTok{$}\NormalTok{values }\OperatorTok{\%\textgreater{}\%}\StringTok{ }\KeywordTok{abs}\NormalTok{()}
\CommentTok{\#\# [1] 4.00 0.25}
\KeywordTok{eigen}\NormalTok{(}\KeywordTok{Re}\NormalTok{(ex\_nonskew\_}\DecValTok{10}\OperatorTok{$}\NormalTok{Theta))}\OperatorTok{$}\NormalTok{values }\OperatorTok{\%\textgreater{}\%}\StringTok{ }\KeywordTok{abs}\NormalTok{()}
\CommentTok{\#\# [1] 4.329881 2.204881}
\end{Highlighting}
\end{Shaded}

\hypertarget{description-of-gmrs-main-procedures}{%
\section{Description of GMR's Main
Procedures}\label{description-of-gmrs-main-procedures}}

\hypertarget{gmm-method}{%
\subsection{GMM Method}\label{gmm-method}}

Let us start by considering the dependency graphs of GMR's GMM method
\emph{estim.VARMAp1.2SLS.GMM(\(\ldots\))}:

\begin{figure}
\centering
\includegraphics{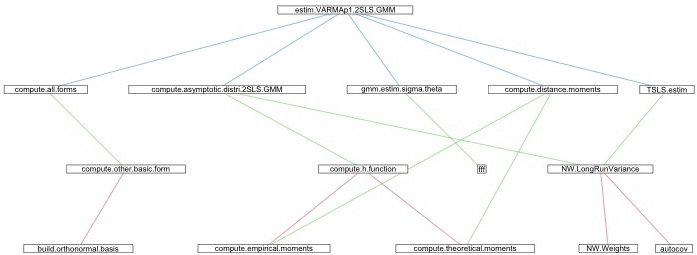}
\caption{Calling tree for GMM procedure}
\end{figure}

The procedure has the following stages.

\begin{enumerate}
\def\labelenumi{\arabic{enumi}.}
\tightlist
\item
  Two stage least squares, implemented in \emph{TSLS(\(\ldots\))}, is
  used to obtain initial estimates for the AR parameters.
\end{enumerate}

\begin{figure}
\centering
\includegraphics{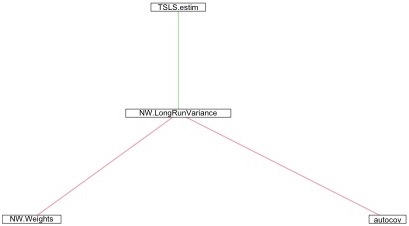}
\caption{Calling tree for two Stage Least Squares}
\end{figure}

\begin{enumerate}
\def\labelenumi{\arabic{enumi}.}
\setcounter{enumi}{1}
\item
  The function \emph{gmm.estim.sigma.theta(\(\ldots\))} is used to
  obtain initial estimates for the MA parameter matrix (and the error
  covariance matrix).
\item
  At this stage the weighting matrix is calculated for the first time
  (with \emph{compute.asymptotic.distri.2SLS.GMM(\(\ldots\))}).
\end{enumerate}

\begin{figure}
\centering
\includegraphics{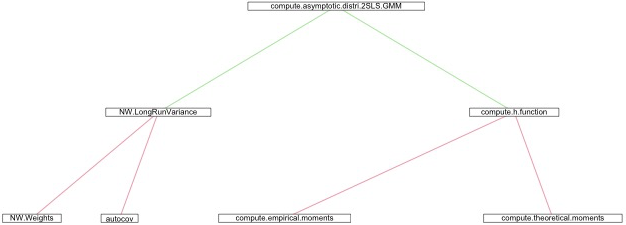}
\caption{Calling tree for computing the asymptotic distribution}
\end{figure}

\begin{enumerate}
\def\labelenumi{\arabic{enumi}.}
\setcounter{enumi}{3}
\tightlist
\item
  The main worker function \emph{compute.distance.moments(\(\ldots\))}
  is called iteratively \emph{nb.loops.GMM = 5} times (a parameter
  specified in the script \emph{run.VARMA.BQ.dataset.R}) by the main GMM
  function \emph{estim.VARMAp1.2SLS.GMM(\(\ldots\))} (which itself is
  called in the script \emph{run.estim.BQ.GMM.MLE.R}) to optimise the
  parameters. In each of the \emph{nb.loops.GMM = 5} iterations,
  \emph{optimx::optimx(\(\ldots\))} is called first with method
  \emph{BFGS} and then with method \emph{Nelder-Mead}. For both methods,
  the maximal number of iterations is set to \emph{maxitNM = 2000}
  (default parameter value, but also in the script
  \emph{run.estim.BQ.GMM.MLE.R}). The name as well as the value of
  \emph{maxitNM} suggests that this is done by accident for the
  \emph{BFGS} method where a different (lower) number of maximal
  iterations should have been chosen (default in R for derivative based
  methods is 100 iterations).
\end{enumerate}

\begin{figure}
\centering
\includegraphics{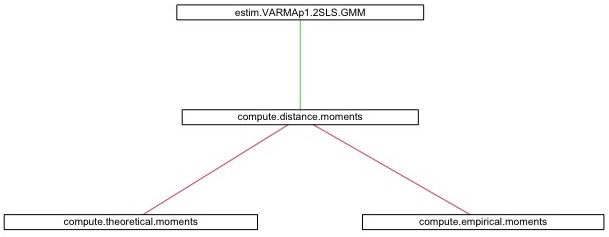}
\caption{Calling tree for computing the distance between theoretical and
empirical moments}
\end{figure}

\begin{enumerate}
\def\labelenumi{\arabic{enumi}.}
\setcounter{enumi}{4}
\item
  Now, all possible Blaschke matrices are calculated. In GMR's empirical
  application there are \(2^2-1=3\) flips. For each of these flips, the
  same procedure is applied (\(5\) iterations of \emph{BFGS} and
  \emph{Nelder-Mead} optimisations). Only if the obtained likelihood
  value is improved, is the ``optimal model'' updated.
\item
  The weighting matrix is calculated a second time (with
  \emph{compute.asymptotic.distri.2SLS.GMM(\(\ldots\))}).
\item
  The \(5\) iterations of \emph{compute.distance.moments(\(\ldots\))}
  are performed one Lastly time with the currently best model.
\item
  The weighting matrix is calculated a third and Lastly time (again with
  \emph{compute.asymptotic.distri.2SLS.GMM(\(\ldots\))}).
\end{enumerate}

\hypertarget{ml-method}{%
\subsection{ML Method}\label{ml-method}}

The optimal parameter from the GMM method serves as initial value for
the ML method. The main work is done by the function
\emph{estim.MA.inversion.aux(\(\ldots\))} (which is called by
\emph{estim.MA.inversion(\(\ldots\))}) The function called/optimised in
\emph{optimx::optimx(\(\ldots\))} below is
\emph{ML.inversion.loglik(\(\ldots\))} (which uses the
Schur-factorization for calculating the residuals).

\begin{figure}
\centering
\includegraphics{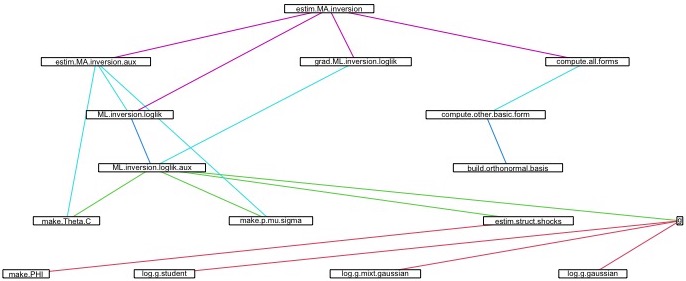}
\caption{Calling tree for ML procedure}
\end{figure}

\begin{enumerate}
\def\labelenumi{\arabic{enumi}.}
\tightlist
\item
  In the first call to \emph{estim.MA.inversion.aux(\(\ldots\))}, the
  optimiser is called \emph{nb.loops.Lastly = 4} times (this time with
  \emph{nlminb} and \emph{Nelder-Mead} with respectively
  \emph{MAXIT.nlminb.Lastly = 300} and \emph{MAXIT.NlMd.Lastly = 1000}
  iterations). These parameters are defined in the script
  \emph{run.VARMA.BQ.dataset.R}.
\end{enumerate}

\begin{figure}
\centering
\includegraphics{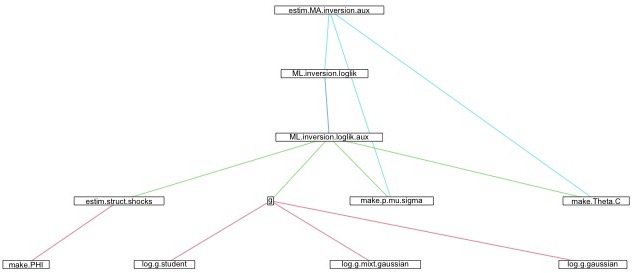}
\caption{Calling tree of main worker for ML procedure}
\end{figure}

\begin{figure}
\centering
\includegraphics{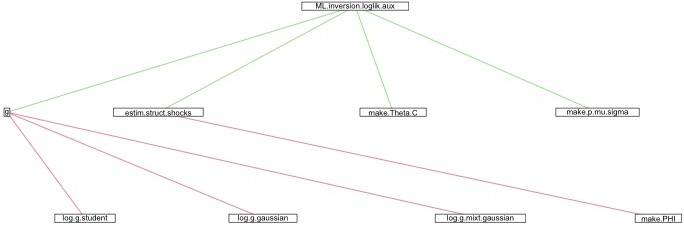}
\caption{Calling tree for likelihood evaluation}
\end{figure}

\begin{enumerate}
\def\labelenumi{\arabic{enumi}.}
\setcounter{enumi}{1}
\item
  Then, all Blaschke flips are obtained and the same optimisation
  procedure is performed but only with \emph{nb.loops.BM = 2} iterations
  for each new starting value obtained by flipping. The fact that the
  Blaschke matrices are complex-valued is covered up at this step.
  Tracking the zero location and the likelihood values is therefore
  particularly interesting at this step. The output of the optimiser is
  saved but the optimum is only updated if there is an improvement.
\item
  Lastly, the covariance is calculated through an optim(\(\ldots\))-call
  to \emph{ML.inversion.loglik(\(\ldots\))} with argument \emph{hessian
  = TRUE}.
\end{enumerate}

\end{document}